\documentclass[sigconf,nonacm]{acmart}

\usepackage{mathrsfs}
\usepackage[title]{appendix}
\usepackage[table]{xcolor}
\usepackage{textcomp}
\usepackage{manyfoot}
\usepackage{tabularx}
\usepackage{booktabs}
\usepackage{algorithm}
\usepackage{algorithmicx}
\usepackage{algpseudocode}
\usepackage{listings}
\usepackage{graphicx}
\usepackage{natbib}
\usepackage{cleveref}
\usepackage{subcaption}
\usepackage{pifont}
\usepackage{verbatim}
\usepackage[breakable]{tcolorbox}
\usepackage{enumitem}
\usepackage{dblfloatfix}

\newcolumntype{C}[1]{>{\centering\arraybackslash}m{#1}}

\makeatletter
\renewcommand{\p@subfigure}{}
\makeatother

\AtBeginDocument{%
  }

\begin{document}

\title{AI-Assisted Social Story Intervention for Special Education: 
The Design of \textit{AdaptED Stories}}

\author{Buyankhishig Enkhjargal}
\email{be2143@nyu.edu}
\affiliation{%
  \institution{New York University Abu Dhabi}
  \city{Abu Dhabi}
  \country{United Arab Emirates}
}

\author{Himanshi Lalwani}
\email{hl3937@nyu.edu}
\affiliation{%
  \institution{New York University Abu Dhabi}
  \city{Abu Dhabi}
  \country{United Arab Emirates}
}

\author{Hanan Salam}
\email{hs4461@nyu.edu}
\affiliation{%
  \institution{New York University Abu Dhabi}
  \city{Abu Dhabi}
  \country{United Arab Emirates}
}

\renewcommand{\shortauthors}{Enkhjargal et al.}

\begin{abstract}
Social Stories are widely used to support autistic children in understanding and preparing for everyday situations, but creating stories that are appropriately tailored to each child’s needs remains labor-intensive for practitioners. Existing digital tools support story assembly and delivery, but much of the work of writing, visual preparation, and personalization remains manual. Recent AI-based approaches have enabled automated story generation, yet offer limited support for practitioner oversight, context-sensitive personalization, and the use of supportive visuals grounded in individual learner profiles. We present \textit{AdaptED Stories}, a practitioner-guided system for authoring, personalizing, and delivering Social Stories in special-education contexts. The system uses student profiles to draft story text and visuals, supports review and refinement by practitioners, and includes reading-session support with comprehension activities and session records. We report findings from a practitioner-informed design process, assessments of generated stories and visuals, and a usability study with seven special-education practitioners. Our findings suggest that AI assistance can reduce story-preparation burden and support more individualized story creation, alongside the importance of practitioner oversight, cultural and contextual specificity, and designing for varied learner needs. These findings contribute design implications for AI-assisted accessibility tools in special-education.
\end{abstract}




\begin{teaserfigure}
    \centering
    \includegraphics[width=\textwidth]{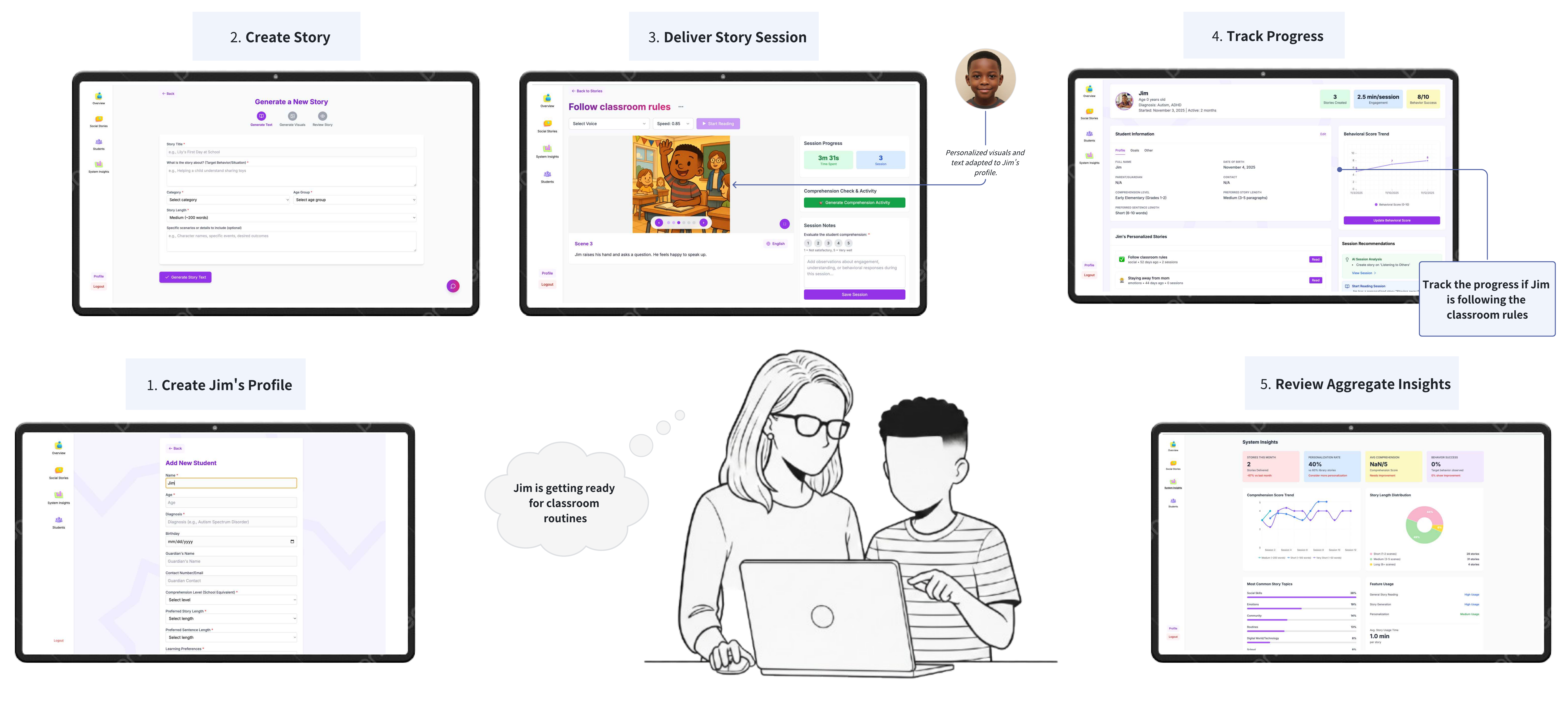}
    \caption{\textit{AdaptED Stories} supports practitioners in creating and delivering personalized Social Stories for autistic children. (1) A student profile captures comprehension level, goals, and preferences. (2) The practitioner generates a story for a target situation, with an LLM drafting text and scene visuals featuring a cartoon avatar of the student. (3) During a reading session, AI narration reads the story aloud while the practitioner records comprehension ratings and session notes. A comprehension activity is generated for the student to practice. (4) The student profile tracks comprehension and goal-related behavioral progress over time. (5) A system-wide Analysis tab aggregates usage and outcome metrics across all students.}
    \label{fig:overview}
\end{teaserfigure}

\maketitle


\section{Introduction}

Autism Spectrum Disorder (ASD) is a neurodevelopmental disorder characterized by persistent challenges in social interaction and communication, alongside restricted and repetitive patterns of behavior, interests, or activities \cite{APA2022}. The prevalence of ASD has been increasingly recognized worldwide \cite{lord2006autism,rice2007public}, with recent estimates suggesting that approximately 1 in 100 children are affected globally \cite{zeidan2022global,who_autism_fact_sheet_2023} and 1 in 31 children aged 8 years (approximately 3.2\%) diagnosed with the condition in the US \cite{cdc_addm_2022}. These challenges often impact children's ability to navigate everyday social situations, adapt to new environments, and form peer relationships, motivating the need for structured and evidence-based interventions that can support social, emotional, and behavioral development.

Among the most widely adopted interventions for supporting children with ASD are Social Stories \cite{wong2015}. Developed by Carol Gray in the early 1990s, these short descriptive narratives are designed to explain unfamiliar social situations and clarify the behavioral expectations associated with them in a structured and reassuring manner \cite{Gray1993, Gray2010}. Social Stories are widely implemented by educational psychologists, teachers, and parents as a tool to prepare children with autism for real-life social scenarios \cite{aldabas2019effectiveness}. Empirical studies suggest that, in school and therapeutic contexts, they can be effective in modifying target behaviors, extending the duration of social engagement, and increasing the frequency of desirable social skills \cite{delano2006effects, karkhaneh2010social}.

A growing body of research suggests that Social Story effectiveness depends heavily on personalization. Studies show that tailoring stories to a child’s developmental and language abilities improves comprehension and behavioral outcomes \cite{kokina2010social, camilleri2024effective}. In fact, Gray's criteria for Social Stories creation emphasize that authors should adopt the learner's perspective and identify information that is directly relevant to the child and the targeted social situation \cite{Gray}. In practice, this requires tailoring narratives to the child's attention span, learning style, interests, and motivations in order to enhance engagement and comprehension \cite{Gray,lofland_socialnarratives}. Personalizing visuals by including photographs, familiar environments, or images of the child can make stories more engaging and help children generalize skills to new situations \cite{edwards2021personalization, ying2016personalised}. Therefore, these findings emphasize that Social Stories are most effective when they are not generic, but rather carefully adapted to the unique developmental profile, interests, and lived contexts of the child.

However, achieving this level of personalization for each child makes Social Story development a highly manual and time-intensive process. Practitioners must carefully adapt narratives to each child's cognitive profile, developmental stage, and specific situational needs, a process that often demands substantial expertise and effort \cite{Wright2016}. To reduce this burden, recent efforts have shifted toward digital formats \cite{SOFAapp2025, AssistiveWarePictello2025} and, more recently, the incorporation of generative artificial intelligence (AI) \cite{ss-gen-dataset}. These systems can automate aspects of the creation process \cite{ssgen2024,EllaKids2025} and have been shown to increase practitioners' confidence in developing and delivering stories \cite{Francese2022}, while also proving feasible for teachers to implement in naturalistic school settings \cite{Smith2020}. Yet they still fall short: existing tools rarely produce truly individualized stories, seldom integrate supportive, customized visuals, and do not provide analytic capabilities to help practitioners monitor comprehension, track behavioral change, or receive recommendations for future interventions. As a result, even with technological advances, Social Story creation continues to lack the level of personalization and data-informed feedback needed to fully support engagement and comprehension at scale, particularly for non-speaking or minimally verbal children \cite{smith2021digitally, vacas2021visual}.

To address these gaps, we present \textit{AdaptED Stories}, an AI-powered web application that supports special-education practitioners in creating and delivering personalized Social Stories for children with ASD. In this study, we investigate the following research question: \textit{How should AI-assisted systems be designed to support practitioners in authoring and delivering personalized Social Stories for children with ASD?}

Figure~\ref{fig:overview} provides an overview of the main components of \textit{AdaptED Stories}, from creating student profiles to generating stories, conducting reading sessions, and reviewing analytics. The application was co-designed with 19 autism practitioners based in the UAE across four phases: Phase~0: an initial scoping phase to identify unmet needs in existing ASD support tools and motivate a focus on Social Stories; Phase~1: formative design studies using early mock-ups to elicit requirements and refine core workflows; Phase~2: iterative prototyping and refinement of a functional web-based system based on practitioner feedback; and Phase~3: a summative usability study to assess the refined system in authentic practice settings. \textit{AdaptED Stories} integrates large language models (LLMs) to generate personalized narratives and complementary visuals from child-specific inputs, and provides tools for story delivery, comprehension tracking, behavioral progress visualization, and AI-driven recommendations.

This work contributes:
\begin{itemize}

    \item \textit{AdaptED Stories}, a practitioner-guided web application for authoring, personalizing, and delivering Social Stories for autistic children in special-education contexts.
    \item Empirical findings on the benefits and limitations of AI assistance for Social Story personalization in practice, based on formative design work and a summative usability study with special-education practitioners.
    \item Design implications for AI-assisted accessibility tools in special-education: personalization, practitioner review, and contextual specificity.

\end{itemize}


\section{Related Work}
\label{related-work}

\subsection{Digital Tools for Social Stories}
To improve accessibility and implementation, Social Stories have shifted towards digital formats. Applications such as Stories Online for Autism \cite{SOFAapp2025}, Pictello \cite{AssistiveWarePictello2025}, and Social Stories Creator and Library \cite{TouchAutismSocialStoriesCreator2025} enable practitioners and families to assemble stories by entering text, uploading photographs, and adding audio or text-to-speech (TTS) narration. Research indicates that digitally mediated stories can enhance parents’ and practitioners’ perceived competence \cite{Camilleri2022}, support use in naturalistic school settings \cite{Smith2020}, and provide a predictable format that supports repeated exposure and child autonomy \cite{smith2021digitally}. This shift has also enabled more advanced solutions, including conversational interfaces \cite{Francese2021,Francese2022}, augmented reality experiences \cite{lee2024applied}, and storytelling robots \cite{vanderborght2012}.

Despite these advances, story creation remains largely manual, as practitioners still write narratives, select or upload images, and adapt each story for individual children. Recent work has applied LLMs to automate parts of this process. SS-GEN uses constraint-driven prompting to generate stories following Carol Gray's guidelines \cite{ssgen2024}, while AutiHero generates story text and illustrations in a parent-facing deployment \cite{lee2025autihero}. However, parent-centered authoring may create tensions between parental priorities and therapeutic goals \cite{lee2025autihero}. This suggests the value of positioning practitioners, who bring formal training and cross-case experience, as the primary authors of therapeutically grounded stories.

More broadly, generative AI has begun to support authoring tasks across special education beyond story generation. Generative AI assistance has been shown to improve the quality and efficiency of writing individualized goals, such as SMART-criteria IEP goals for preschoolers with autism \cite{rakap2024chatting}. Multi-agent LLM systems have similarly been proposed to generate personalized classroom worksheets, though single-LLM approaches have been found to lack the pedagogical depth needed for comprehensive learner modeling \cite{gonnermann2025facet}. These efforts reflect a broader shift toward AI-assisted authoring in special education, motivating our focus on practitioner-facing tools that keep clinical judgment central to the authoring process.

Practitioners draw on formal training and experience across multiple children, allowing them to select target situations, frame behaviors, and interpret story effects in ways that align with evidence-based practice \cite{hugh2024preschool, layden2024discovering}. Yet existing digital tools provide limited support for practitioner-led use, child-specific visual personalization, and integrated monitoring of comprehension and behavior. \textit{AdaptED Stories} responds to these gaps through a practitioner-facing system for guided story authoring, personalization, review, and follow-up support.

\subsection{Social Story Design and Personalization}
Personalization is a defining feature of Social Stories. Carol Gray's guidelines for Social Story creation explicitly emphasize this requirement \cite{Gray}. Criterion~2 (Discovery: Story or Alternate Solution?) directs authors to identify relevant details about the child and the social context before writing, while Criterion~4 (Format—Tailor \& Personalize) requires that each story be adjusted to the learner's ``abilities, attention span, learning style and, whenever possible, talents and interest'' \cite{Gray}. Research further indicates that personalization must also have a cultural dimension as social narratives developed in Western contexts may miss locally meaningful cues and expectations \cite{khan2024recognizing, tawankanjanachot2023systematic}. Practitioner-oriented online resources similarly note that tailoring stories to a child’s age, comprehension level, interests, and everyday experiences maximizes their effectiveness \cite{AdinaABA2025, IntelliStarsABA2025}.

Research also emphasizes the importance of visual personalization. Edwards et al.\ compared delivery formats and found that incorporating personalized images, or combining them with role play, led to stronger improvements in target behaviors than generic deliveries \cite{edwards2021personalization}. Chen et al. developed a video face-replacement system that embedded familiar faces and environments into social narratives; children exposed to these personalized videos showed more positive affect during novel experiences \cite{7349713}. Similarly, combining Social Stories with video self-modeling has been shown to improve target social behaviors such as greeting and initiating play, with skills transferring across contexts \cite{litras2010using}. 

Collectively, these findings demonstrate that personalizing both text and visuals enhances engagement, comprehension, and generalization. However, prior approaches often rely on manual preparation or specialized technologies, which limits their scalability and ease of use in everyday practice \cite{edwards2021personalization, 7349713}. Building on this evidence, \textit{AdaptED Stories} supports AI-assisted personalization of both story text and visuals while reducing the manual effort required to adapt stories for individual learners.

\begin{figure}[h]
  \centering
  \includegraphics[width=0.5\textwidth]{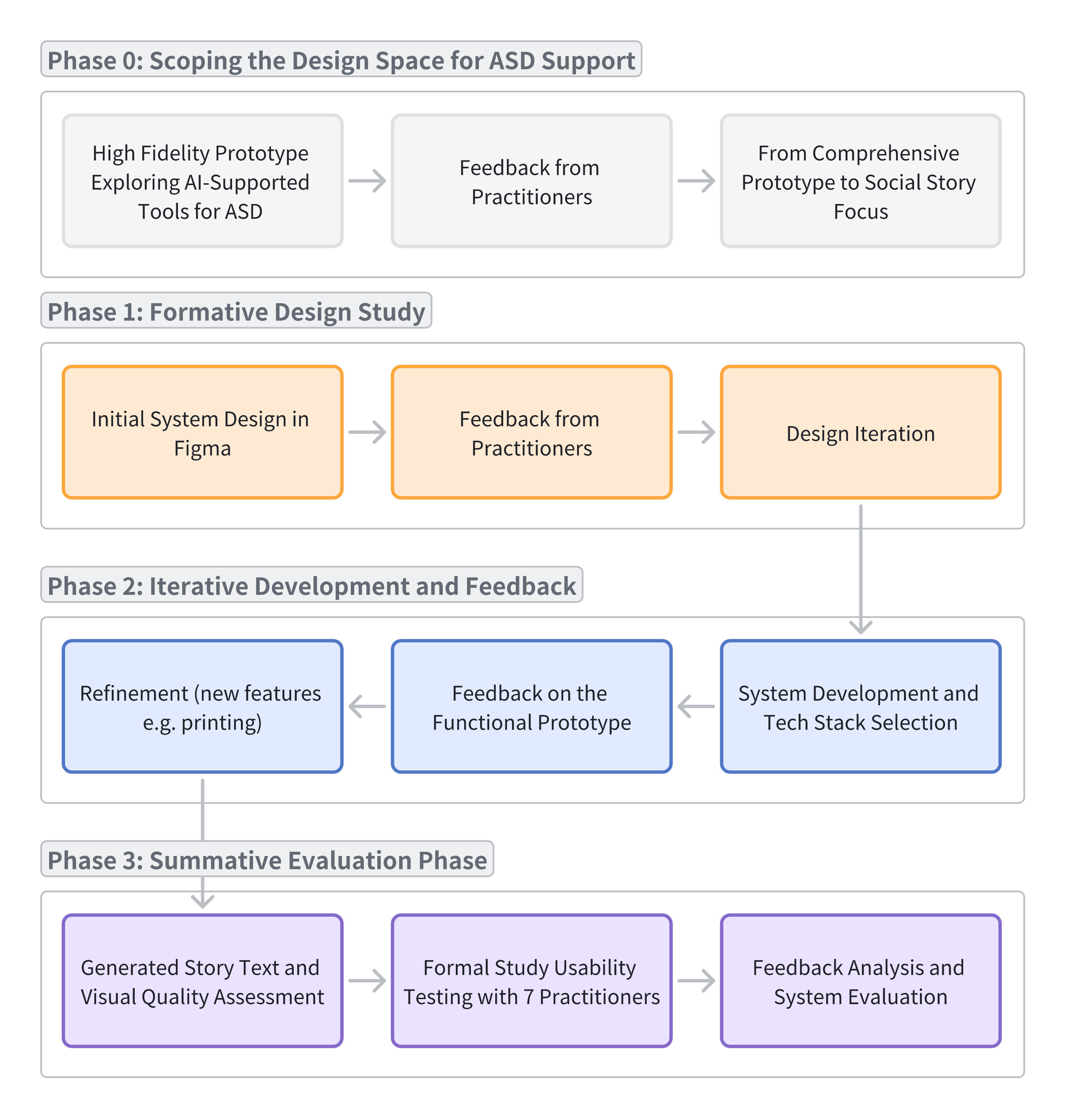}
  \caption{Overview of the four-phase, practitioner-informed methodology. Phase~0 scoped the design space using a high-fidelity multi-tool prototype for ASD support, leading us to focus on Social Stories. Phase~1 translated this focus into Figma mock-ups and refined core features through formative design sessions with practitioners. In Phase~2, we implemented a functional web-based \emph{AdaptED Stories} prototype and iteratively improved it based on practitioner feedback. Phase~3 used the refined system for systematic assessment of story text and visuals and for a summative usability study with seven practitioners.}
  \label{fig:phases}
\end{figure}

\subsection{Practitioner-Informed Design}

Participatory design traditions in HCI emphasize involving relevant stakeholders in shaping technologies intended for their use \cite{liegl2016designing, bjogvinsson2012design, bodker2022participatory}. Prior work has applied such approaches in the design of technologies for autistic children and other people with special needs, including systems co-designed with autistic youth and tools developed with domain stakeholders to support communication and learning \cite{fabri2016using, parsons2011participatory, 10.1145/3613904.3642080}. Such approaches are valuable because they help align resulting systems with stakeholders’ practices, needs, and constraints, while also contributing intangible benefits such as new practices, procedures, and shared knowledge \cite{bodker2022participatory, hansen2019participatory}.

In our case, the system is intended to support practitioners in authoring, personalizing, and delivering Social Stories. Because Social Stories are used by a wide range of professionals who work with autistic children, we define practitioners broadly as individuals who author and deliver these interventions. In this study, we included practitioners from several such backgrounds, including speech-language pathologists (SLPs), ABA therapists, special education teachers, and program directors. Given this range of roles, our work follows a practitioner-informed design process in which autism practitioners contributed by identifying unmet needs in existing tools, reacting to early concepts and mock-ups, refining feature requirements and workflows, and evaluating successive prototypes. The research team synthesized this input, translated it into design requirements and implemented the system. 


\begin{figure*}[ht]
    \centering
    \begin{subfigure}[b]{0.49\textwidth}
        \centering
         \captionsetup{justification=raggedright, singlelinecheck=false, labelfont=bf, labelsep=period, font=small} 
        \caption{}
        \includegraphics[width=\textwidth]{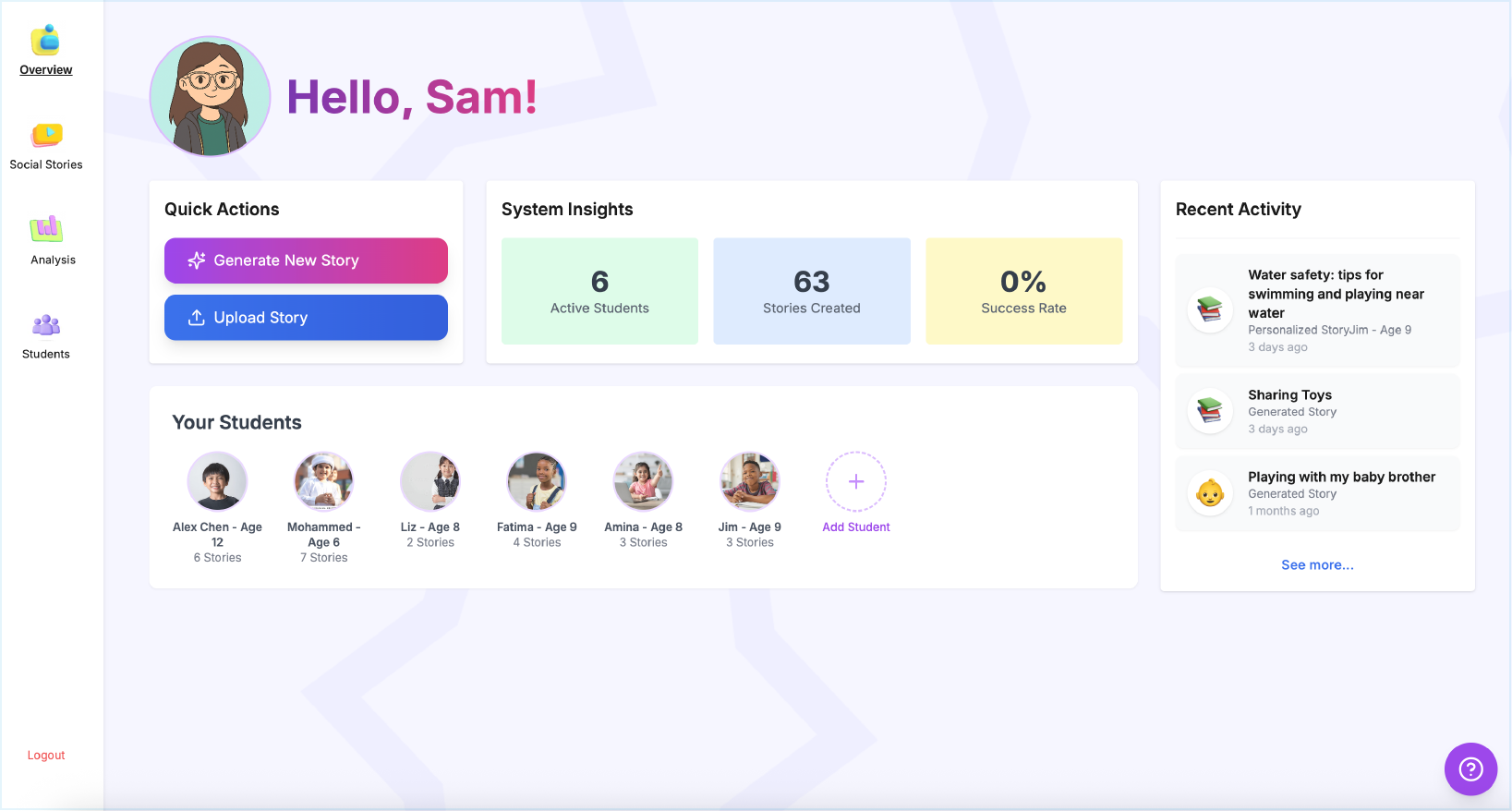} 
        \label{fig:final-overview}
    \end{subfigure}
    \hfill
    \begin{subfigure}[b]{0.49\textwidth}
        \centering
         \captionsetup{justification=raggedright, singlelinecheck=false, labelfont=bf, labelsep=period, font=small} 
        \caption{}
        \includegraphics[width=\textwidth]{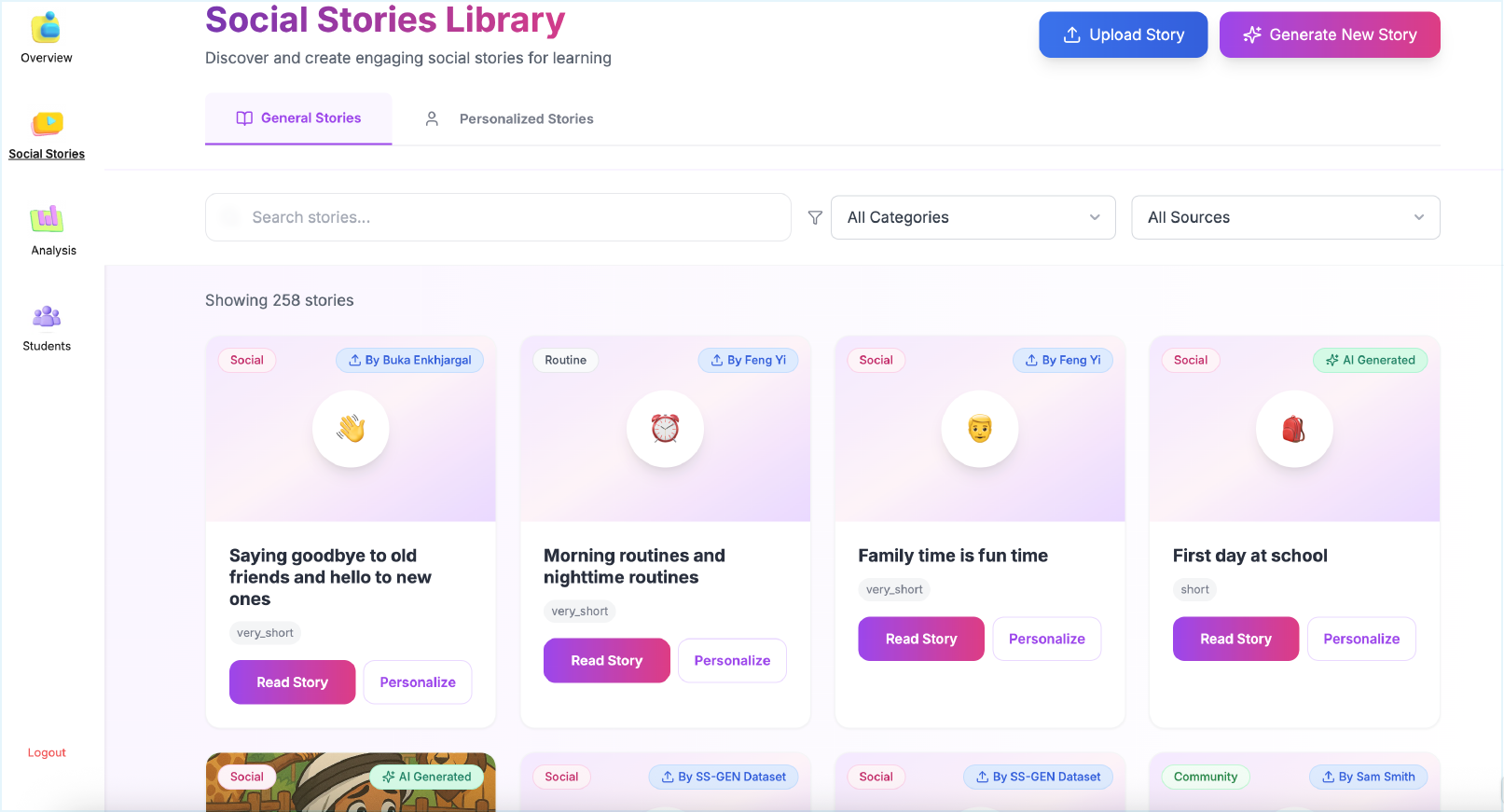} 
        \label{fig:final-ss}
    \end{subfigure}

    \vspace{1em} 

    \begin{subfigure}[b]{0.49\textwidth}
        \centering
         \captionsetup{justification=raggedright, singlelinecheck=false, labelfont=bf, labelsep=period, font=small} 
        \caption{}
        \includegraphics[width=\textwidth]{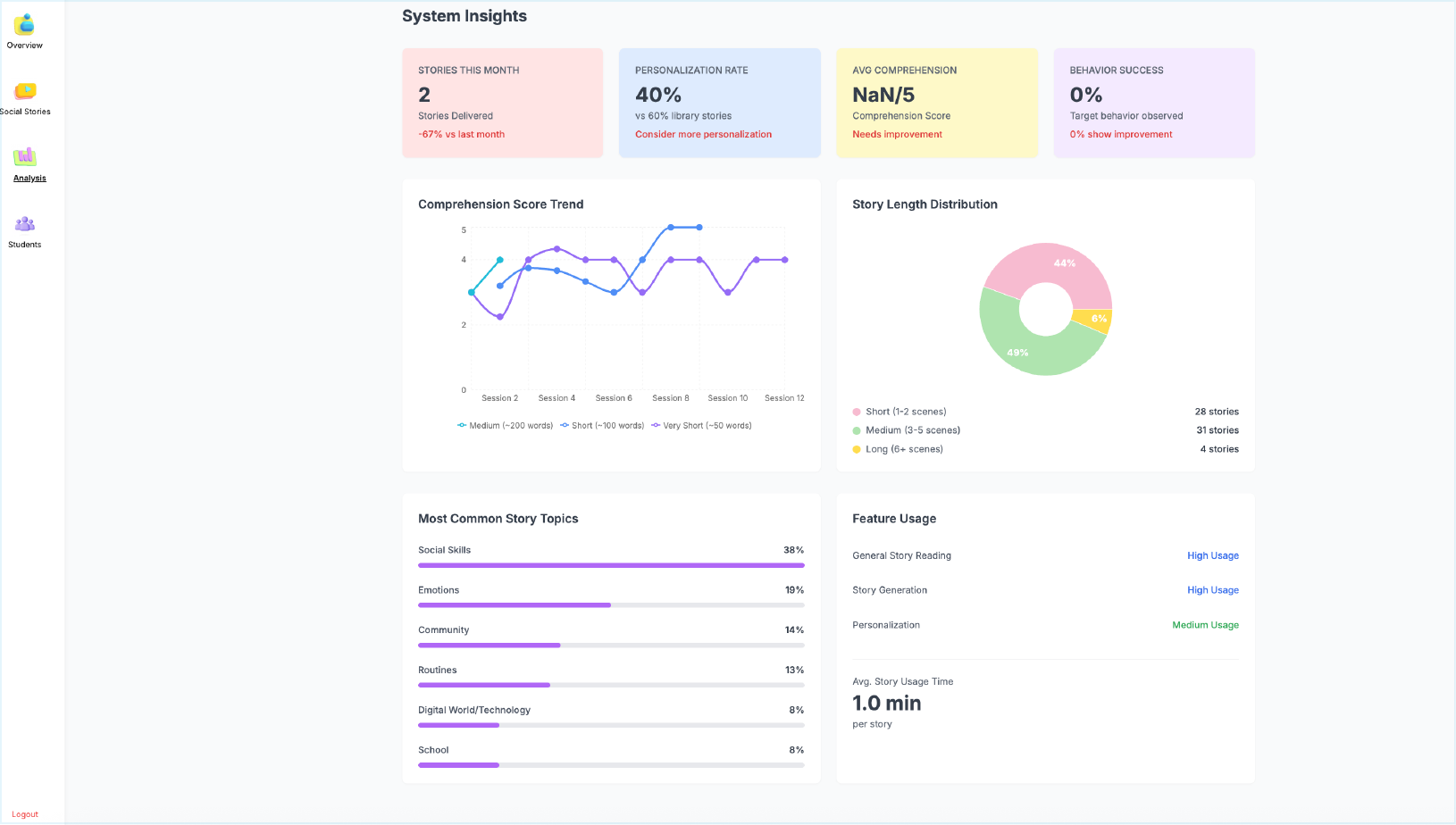} 
        \label{fig:final-analysis}
    \end{subfigure}
    \hfill
    \begin{subfigure}[b]{0.49\textwidth}
        \centering
         \captionsetup{justification=raggedright, singlelinecheck=false, labelfont=bf, labelsep=period, font=small} 
        \caption{}
        \includegraphics[width=\textwidth]{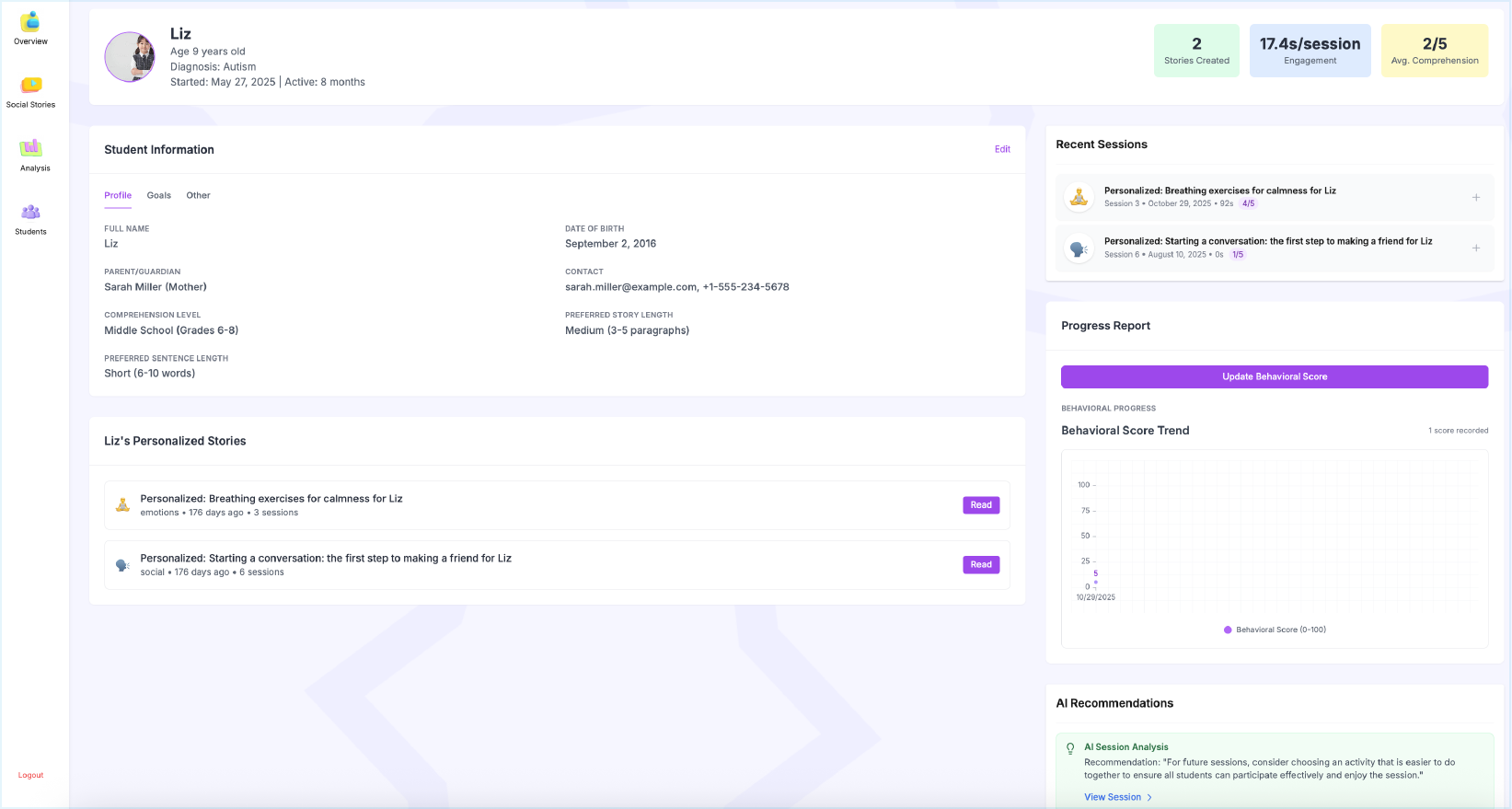} 
        \label{fig:final-students}
    \end{subfigure}

    \caption{Interface views from the final \textit{AdaptED Stories} system. (A) Overview tab showing quick actions and recent activity for the practitioner. (B) Social Stories tab with access to the system library, personalized stories, and the option to create stories using AI. (C) Analysis tab showing student-wide aggregate patterns and system usage analytics, including comprehension trends and story use. (D) Students tab showing an individual student profile with key information, personalized stories, progress summaries, and AI-generated recommendations.}
    \label{fig:final-mockups}
\end{figure*}

\section{Methodology Overview}
\label{sec:methodology}

Our research followed a four-phase, practitioner-informed design and evaluation process (Figure~\ref{fig:phases}). The study protocol was approved by the University Institutional Review Board. Although the work unfolded chronologically from Phase~0 to Phase~3, in the following sections we first present the final \emph{AdaptED Stories} system to ground the reader (Section \ref{sec:system-overview}), and then describe how each phase shaped its design (Section~\ref{sec:design-process}) and evaluation (Section~\ref{sec:evaluation}).

Phase~0 was an exploratory scoping phase in which practitioners reviewed a high-fidelity, AI-enhanced prototype for autism support. Their feedback directed to Social Stories as the area of greatest perceived need and informed our decision to focus subsequent phases on Social Story interventions.

Phase~1 translated this focus into concrete interaction concepts and requirements. Using early Figma mock-ups of an AI-supported Social Story platform, we conducted design sessions with practitioners to refine the information architecture and key workflows for student profiles, story generation and personalization, visual supports, and delivery.

Phase~2 centered on iterative system development and prototype testing. We implemented a functional web-based prototype of \emph{AdaptED Stories} and gathered practitioner feedback on workflow usability and generated content, using these insights to refine the interface, prompts, and session-support features.

Phase~3 consisted of a summative evaluation with special-education practitioners. This phase combined a structured assessment of generated story text and visuals with usability testing of the refined system, focusing on ease of use, intent to adopt, perceived workload reduction, and support for practice. Across all phases, practitioners working with autistic children directly informed the system’s development, helping align \emph{AdaptED Stories} with professional expertise and real-world constraints.

\section{AdaptED Stories: System Overview}
\label{sec:system-overview}
\textit{AdaptED Stories} is a practitioner-facing system organized around four tabs (Figure~\ref{fig:final-mockups}): Overview, which summarizes recent activity and quick actions; Social Stories, for accessing, generating, and personalizing stories; Students, for managing student profiles and reviewing individual progress; and Analysis, for system-wide analytics. An interactive onboarding tour introduces practitioners to these workflows on first sign-in. 

\subsection{Design Goals}
Building on the Social Stories literature and our review of existing digital systems (cf. Section \ref{related-work}), we identified three design goals for \textit{AdaptED Stories}.

\begin{enumerate}
    \item \textbf{DG1: Keep practitioners in control while using AI to reduce authoring effort.} Practitioners bring formal training in behavioral assessment and intervention design that shapes how Social Stories are selected, framed, and used in practice \cite{hugh2024preschool,layden2024discovering}, yet existing AI-assisted systems provide limited support for professional review of generated content before use with children \cite{ssgen2024,lee2025autihero}. Our first goal was to keep practitioners as the main authors and decision-makers, with AI used to draft story text and visuals that are always subject to review and editing. This preserves professional oversight and alignment with evidence-based practices while still reducing the time and effort required to create stories.

    \item \textbf{DG2: Support rich personalization of both story text and visuals.} Guidelines for Social Story creation emphasize that stories should be tailored to a child’s abilities, attention span, interests, and everyday contexts \cite{Gray, AdinaABA2025}. Research further shows that personalized visuals, including photographs or familiar environments and characters, can increase engagement and help children generalize skills across situations \cite{7349713, edwards2021personalization}. Our second goal was therefore to support fine-grained control over story complexity and to allow children to appear as protagonists in story visuals, guided by practitioner-defined student profiles.
    
    \item \textbf{DG3: Connect story delivery with analytics and recommendations.} Existing digital Social Story applications typically focus on authoring, with no support for tracking how children respond to stories over time or for using that information to inform future interventions \cite{TouchAutismSocialStoriesCreator2025, AssistiveWarePictello2025}. As a result, practitioners often rely on informal paper-based notes when monitoring comprehension and behavioral change, which are difficult to aggregate and revisit systematically \cite{marcu2020collaborative, marcu2013they}. Our third goal was to extend beyond story creation by integrating mechanisms for recording comprehension and behavior during reading sessions, summarizing these trends over time, and using session data to generate recommendations that can inform subsequent story selection and personalization. Keeping these records and suggestions within the same application alongside authoring and delivery was intended to reduce application switching and documentation burden and help practitioners make more data-informed choices about what to implement next.
\end{enumerate}

We next describe how these design goals are realized in the implemented system.

\begin{figure}[h]
  \centering
  \includegraphics[scale=0.2]{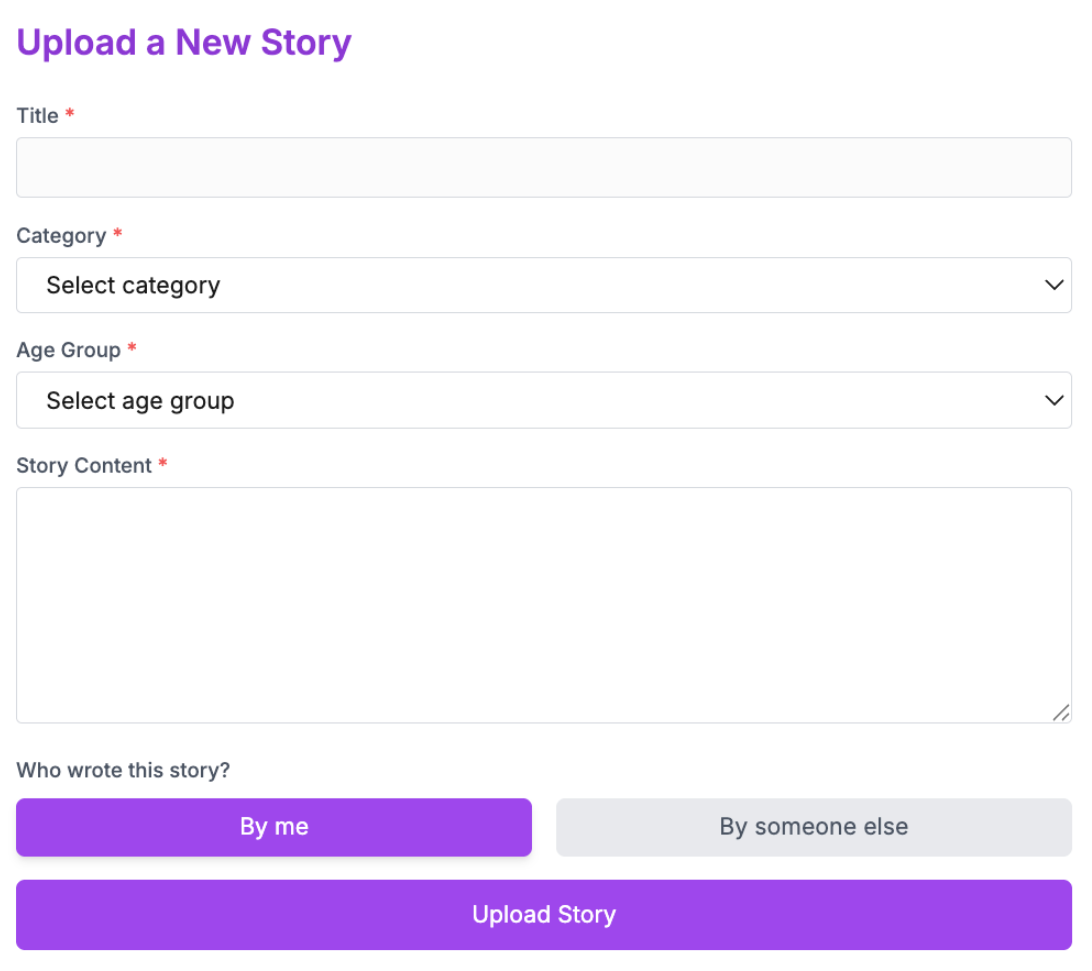}
  \caption{Story upload interface. Practitioners can add existing Social Stories to the system by entering the title, category, age group, and story text, and by indicating who authored the story for proper attribution.}
  \label{fig:upload-story}
\end{figure}

\begin{figure*}
    \centering
    \includegraphics[width=1\linewidth]{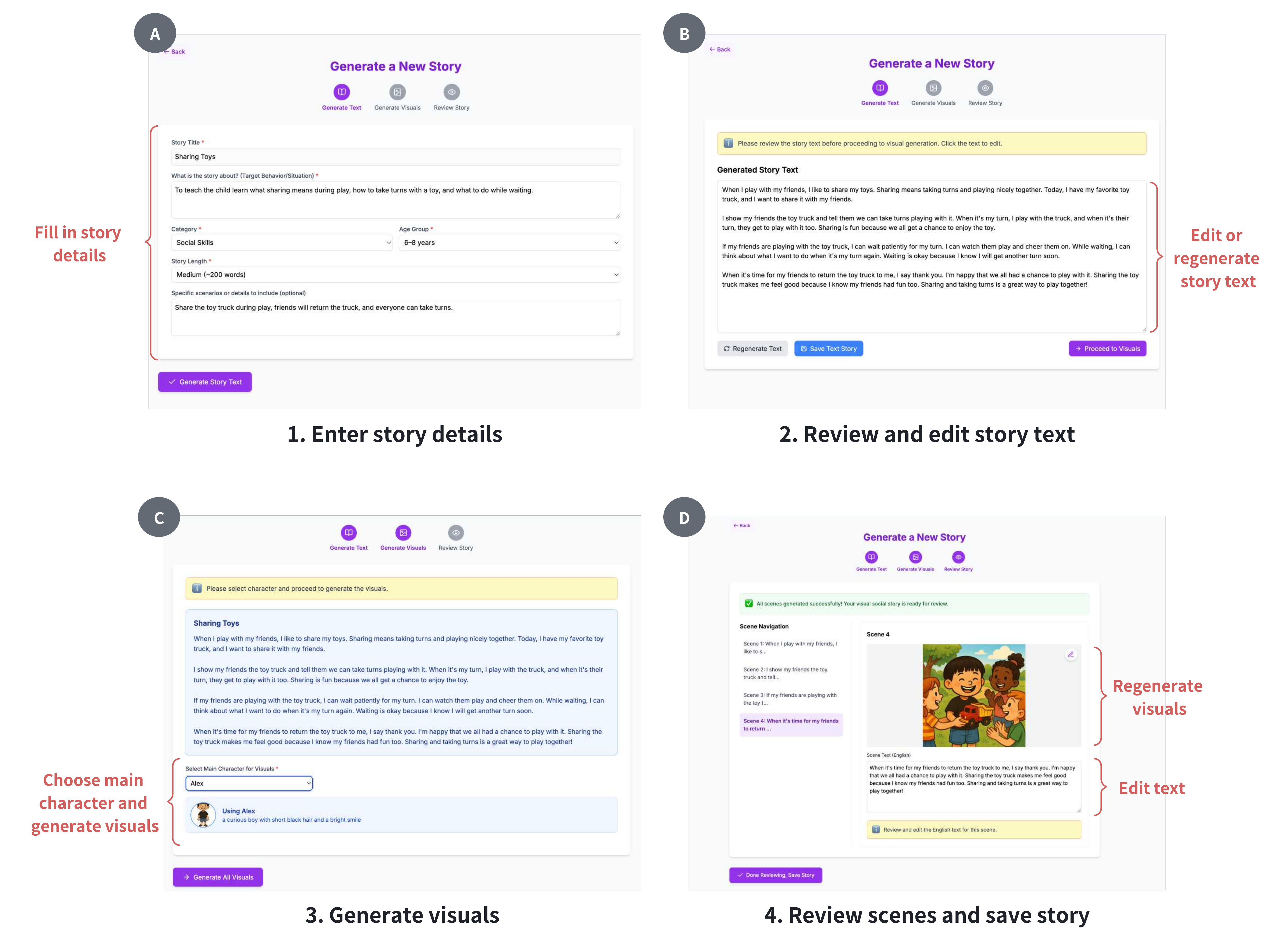}
    \caption{Story creation flow.
(A) The practitioner enters the story title, target situation, category, age group, and length.
(B) The system generates draft story text, which the practitioner can review, edit, or regenerate.
(C) The practitioner selects a base character and requests AI-generated scene visuals.
(D) For each scene, they can adjust the text, regenerate individual visuals if needed, and then save the completed story.}
    \label{fig:story-creation-mockup}
\end{figure*}

\begin{figure*}
        \centering
        \includegraphics[width=1\linewidth]{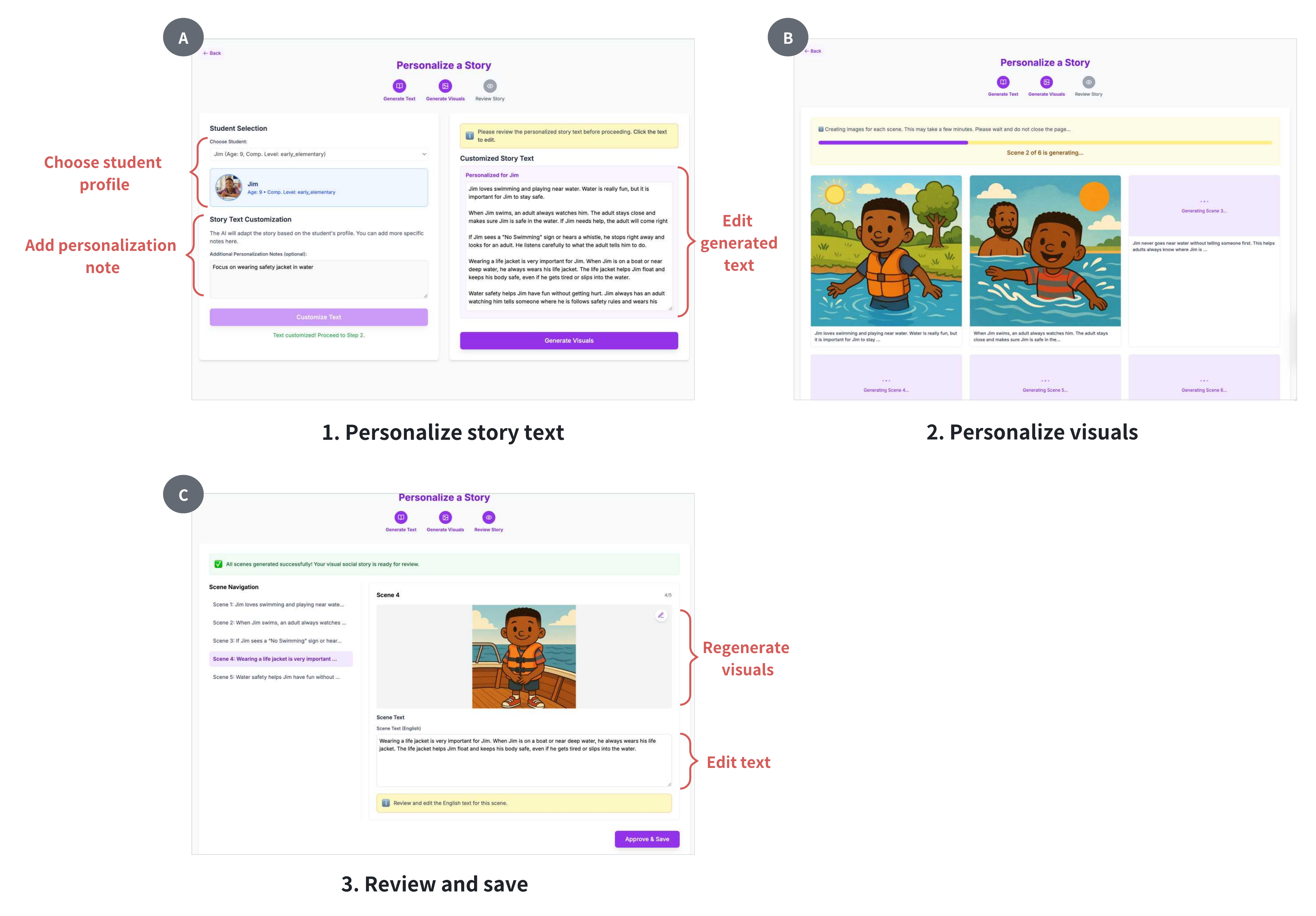}
        \caption{Personalizing an existing Social Story for a specific student. (A) The practitioner selects a student profile and adds an optional personalization note, then reviews and edits the AI-customized story text. (B) The system generates scene-level visuals that the practitioner reviews for relevance and appropriateness for the student. (C) The practitioner reviews each scene, optionally regenerates visuals or refines text, and then approves and saves the personalized story.}
        \label{fig:personalization-mockup}
    \end{figure*}

  \begin{figure*}
        \centering
        \includegraphics[width=1\linewidth]{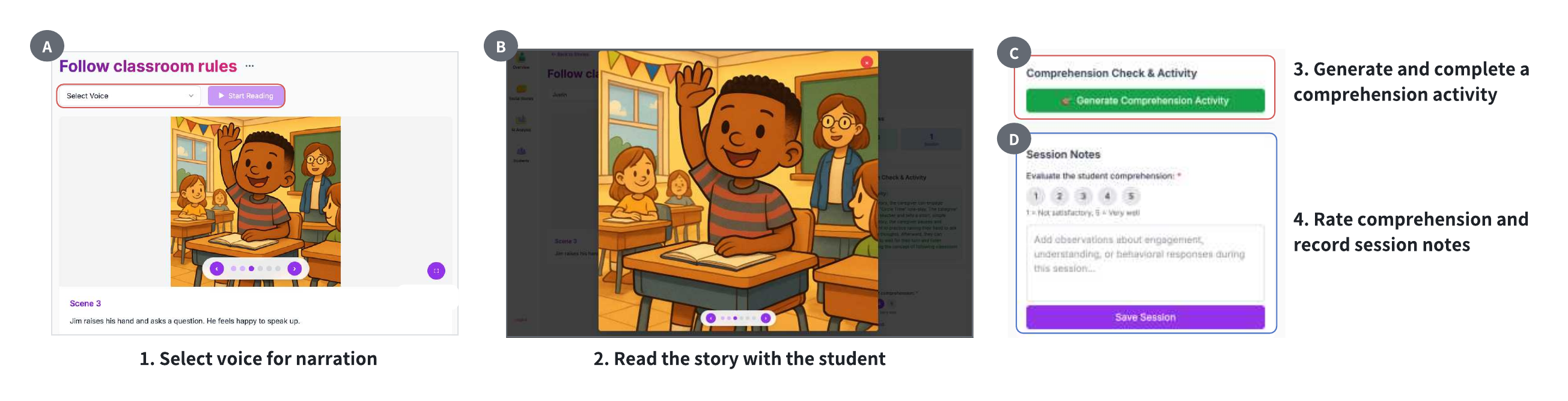}
        \caption{Reading session interface. (A) Practitioner selects a narration voice and starts the story. (B) The story is read with the student using the visual viewer. (C) A single AI-generated comprehension activity replaces earlier multiple-choice questions. (D) The practitioner records a comprehension rating and notes, which are saved with the session.}
        \label{fig:refined-reading-story}
    \end{figure*}

\subsection{Student Profiles}
To personalize stories, practitioners first create a student profile capturing key identifying information, such as the student's name, age, date of birth, diagnosis, and parent or guardian details, alongside attributes used directly for personalization. These include a grade-based comprehension level ranging from Pre-K through Postsecondary, preferred story length ranging from very short (around 5 sentences) to long (more than 5 paragraphs), preferred sentence length ranging from very short (around 5 words) to long (more than 15 words), learning goals and target behaviors, interests, challenges, and a student photograph. The profile serves as the foundation for story generation, personalization, and AI-generated recommendations throughout the system. In this way, student profiles operationalize DG2 by grounding personalization in practitioner-defined information about the learner.

\subsection{Story Library, Generation, and Personalization}
\textit{AdaptED Stories} maintains a shared story library that supports both reuse and new authoring. The library is pre-loaded with 200 validated Social Stories from the SS-GEN dataset \cite{ssgen2024, ss-gen-dataset}, covering six thematic categories: social skills, routines, emotions, community, technology, and school. Practitioners can browse these stories for direct use, personalize them for individual students, generate new stories from scratch, or upload existing stories they already use in practice.

The Social Stories tab is organized around two main views: \textit{General Stories} and \textit{Personalized Stories}. \textit{General Stories} contains the shared library, including pre-loaded stories, practitioner-uploaded stories, and stories generated through the system. \textit{Personalized Stories} contains versions of stories tailored to individual students and saved for repeated use. Stories generated or uploaded by any practitioner are added back to the shared library, enabling it to grow collectively, while personalized stories remain visible only to the practitioner who created them in order to protect student privacy. The library also supports keyword search and filtering by category, student, and story source, enabling practitioners to quickly retrieve stories during everyday school or therapy routines.

Practitioners can also upload existing stories through a dedicated form (see Figure~\ref{fig:upload-story}) that captures the story title, category, age group, full story text, and authorship information for copyright attribution. This supports continuity with practitioners’ existing materials while allowing those stories to be incorporated into the same authoring and delivery environment.

To generate a new story (Figure~\ref{fig:story-creation-mockup}), practitioners specify the story title, category, age group, target situation, story length, and preferred sentence length. The system then produces a draft narrative that practitioners review, edit, regenerate if needed, and explicitly approve before proceeding to visual generation. This staged workflow separates text approval from visual approval so that practitioners evaluate each modality independently rather than working with a bundled output. After confirming the text, practitioners select one of five pre-defined cartoon characters representing diverse race and gender combinations, and the system generates scene-level visuals. Each scene can be individually reviewed and regenerated before the completed story is saved to the library.

To personalize an existing story for a specific student (Figure~\ref{fig:personalization-mockup}), practitioners select a story from the library and choose a student profile, with an optional free-text field for additional personalization notes. The system generates adapted story text informed by the student profile, which practitioners review and edit before confirming. Once the text is approved, the system generates scene-level visuals.  For personalized stories, the visuals are generated using the student's cartoon avatar from their profile. To support visual consistency across scenes, a base cartoon image of the child is generated once, at profile creation, and then reused as a character reference whenever personalized visuals are produced. This reference helps preserve attributes such as skin tone, hairstyle, and facial features, so the student consistently appears as the protagonist. Background inconsistencies may still occur. Once all visuals have been generated, practitioners can inspect each scene, regenerate individual visuals, and make further text edits before saving the personalized story to that student’s profile.

Across both generation and personalization workflows, AI outputs are treated as reviewable drafts rather than finished products. Together, these workflows realize DG1 and DG2 by combining AI-assisted drafting with explicit practitioner review and student-specific personalization.

\subsection{Reading Sessions and Comprehension Activities}
Stories can be launched from either the shared library or a student’s personalized story collection. During a reading session (see Figure~\ref{fig:refined-reading-story}), the story is automatically presented one scene at a time with its associated visual and text-to-speech narration. Practitioners can choose between two child-like voices, one male and one female, and read along with the child as the narration plays.

\begin{figure}[h]
  \centering
  \includegraphics[scale=0.6]{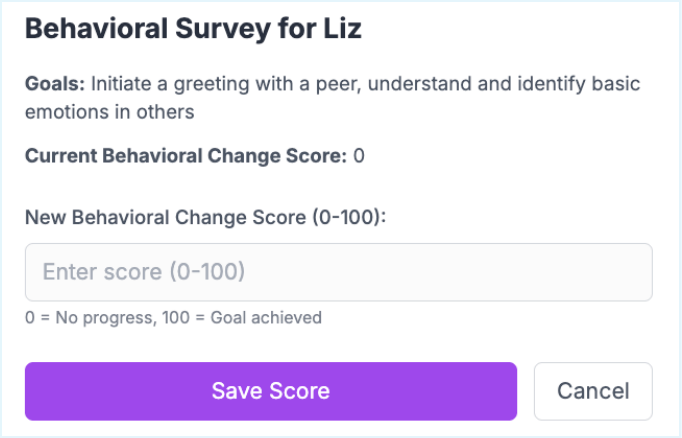}
  \caption{Behavioral survey displaying the student’s goal, current behavioral score, and an input field for entering a new score. This survey tracks behavioral changes during the social story intervention period. It is presented to the practitioner for each story reading session to track student target behavior development.}
  \label{fig:behaviour-survey}
\end{figure}

Before each session, the system displays a brief behavioral survey (see Figure~\ref{fig:behaviour-survey}) linked to the student’s current goal, prompting the practitioner to enter an updated behavioral change score on a 0-100 scale. This score is recorded as part of the student’s history and can be skipped if not relevant to the session.

After the story is read, the system uses an LLM to generate a short interactive comprehension activity from the story content. Designed to be completed on the spot in roughly 2-3 minutes, the activity is intended to help the child practice and apply the story content rather than answer recall-based questions. Practitioners then record a comprehension rating on a 1-5 scale that captures their perception of how well the child understood the story in that session, together with optional free-text session notes describing the child's engagement and behavioral responses. Stories can also be downloaded as PDFs for offline use or sharing with families. The system logs comprehension scores, behavioral survey scores, session duration, and notes as part of the student's history.

\begin{figure*}
  \centering  \includegraphics[scale=0.2]{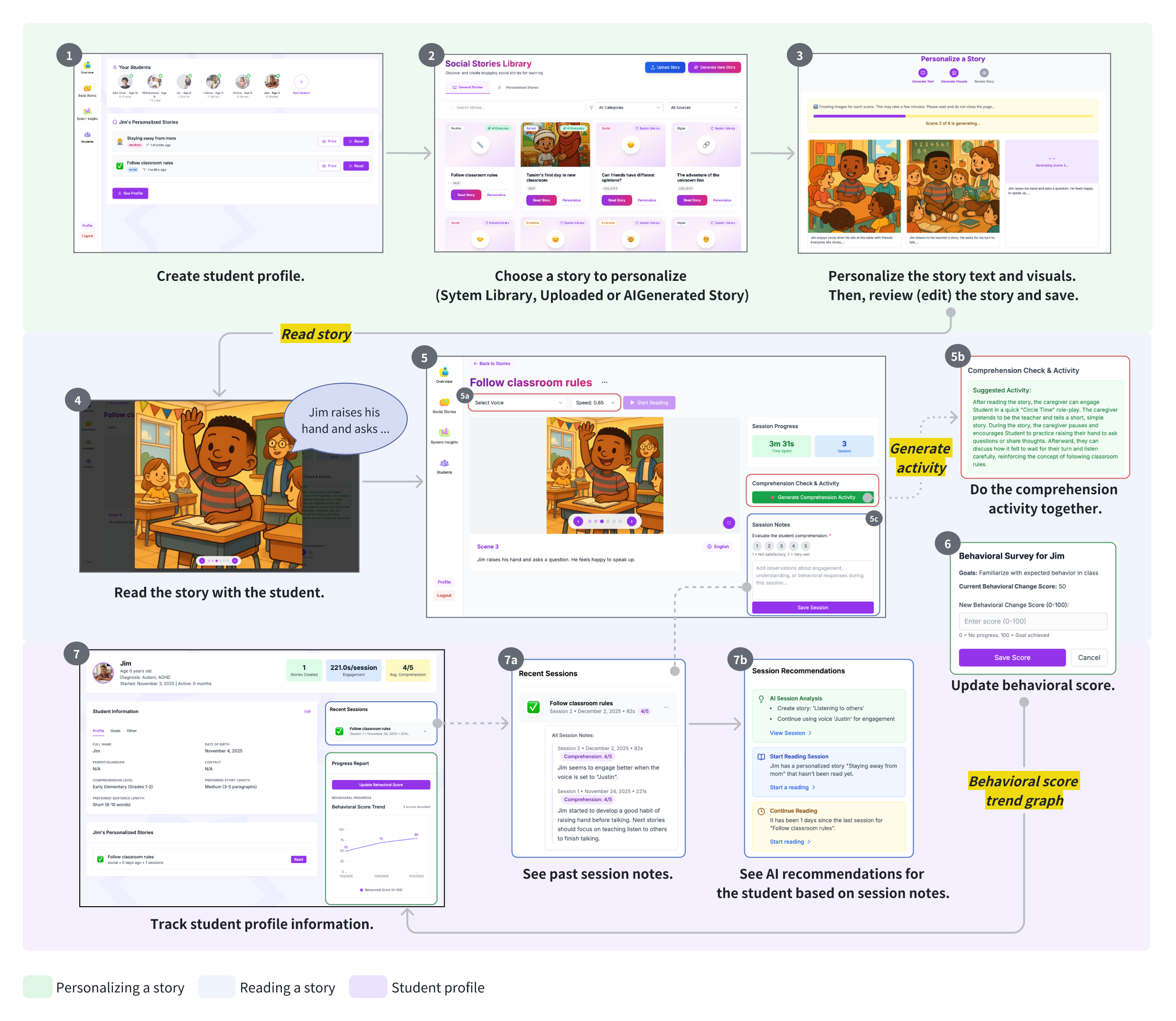}
  \caption{Main screens and usage flow of \textit{AdaptED Stories}: (1) Students tab - Practitioners can view students, access personalized stories, and create new profiles. (2) Social Stories tab - Practitioners can select a story and click the ``personalize'' button to begin personalizing it. (3) Story personalization process - A three-step process covering text personalization, visual personalization, and final review/editing before saving. (4) The saved story is then read on the (5) Reading Page, where practitioners first (5a) choose a voice, can also (5b) generate a comprehension-check activity, and finally (5c) record a comprehension score and optional session notes. Session notes appear on the student’s (7) Profile Page, where (7a) all notes are used to generate (7b) AI recommendations for future social story interventions. Additionally, each reading session includes an optional (6) behavioral pop-up survey, and score trends from these surveys are displayed as a graph on the (7) Profile Page..}
  \label{fig:user-interaction}
\end{figure*}

These workflows connect story authoring and personalization with story delivery and immediate follow-up, directly supporting DG3. Figure~\ref{fig:user-interaction} summarizes this end-to-end interaction sequence, from creating a student profile and personalizing a story to reading it with the child and recording session outcomes.

\subsection{Session Records, Analytics, and Recommendations}
Across reading sessions, \textit{AdaptED Stories} aggregates comprehension ratings, behavioral scores, session duration, and session notes to support practitioner reflection and planning.

At the level of an individual student, the profile page in the Students tab presents longitudinal views of comprehension and behavior scores alongside a history of session notes. Practitioners can review how a child has responded to stories over time and relate those trends to their qualitative observations. Session notes are also processed by the system to generate concise, student-specific recommendations for future story selection and personalization, such as revisiting particular topics or adjusting story complexity.

At the cohort level, the Analysis tab provides an aggregate view of system use and outcomes across students. Summary metrics include stories delivered, personalization rate, average comprehension score, and feature usage patterns, accompanied by visualizations of comprehension trends, common story topics, and story-use patterns. These views are intended to support broader planning and provide a complementary perspective to the student-level records.

\subsection{Technical Implementation}
\label{sec:technical-implementation}
\paragraph{\textbf{Application stack}}
\textit{AdaptED Stories} is implemented as a web application using Next.js, with MongoDB for data storage, Redis for caching, and Cloudinary for image storage.

\paragraph{\textbf{AI infrastructure}}
Story text generation, personalization, comprehension 
activities, and AI recommendations use Google 
Gemini 2.0 Flash~\cite{geminiflash}, selected based on a 
comparative study finding a favorable balance of relevance, 
consistency, tone, and latency for storytelling 
applications~\cite{LLMrobot2025}. For image generation, we 
integrate the GPT-4o~\cite{gpt4o} API to generate cartoon-style 
scene visuals from the student's profile photograph and story 
text. The research team selected GPT-4o after comparing its output quality and cartoon style adherence against Imagen 3 for this use case. Prompt templates are provided in 
Appendix~\ref{sec:prompts}. For narration, we use Amazon 
Polly~\cite{awsPolly} with child voices Ivy and 
Justin~\cite{pollyVoices}, and the SSML tag support enables speech 
rate adjustment to match different comprehension levels.

\section{Practitioner-Informed Design Process}
\label{sec:design-process}
Building on the overview in Section~\ref{sec:methodology}, we now describe Phases~0-2 of our practitioner-informed design process in more detail, showing how practitioner input first motivated the idea of \textit{AdaptED Stories} and then informed key design decisions in the system.

\begin{figure}[ht]
  \centering
  \includegraphics[scale=0.22]{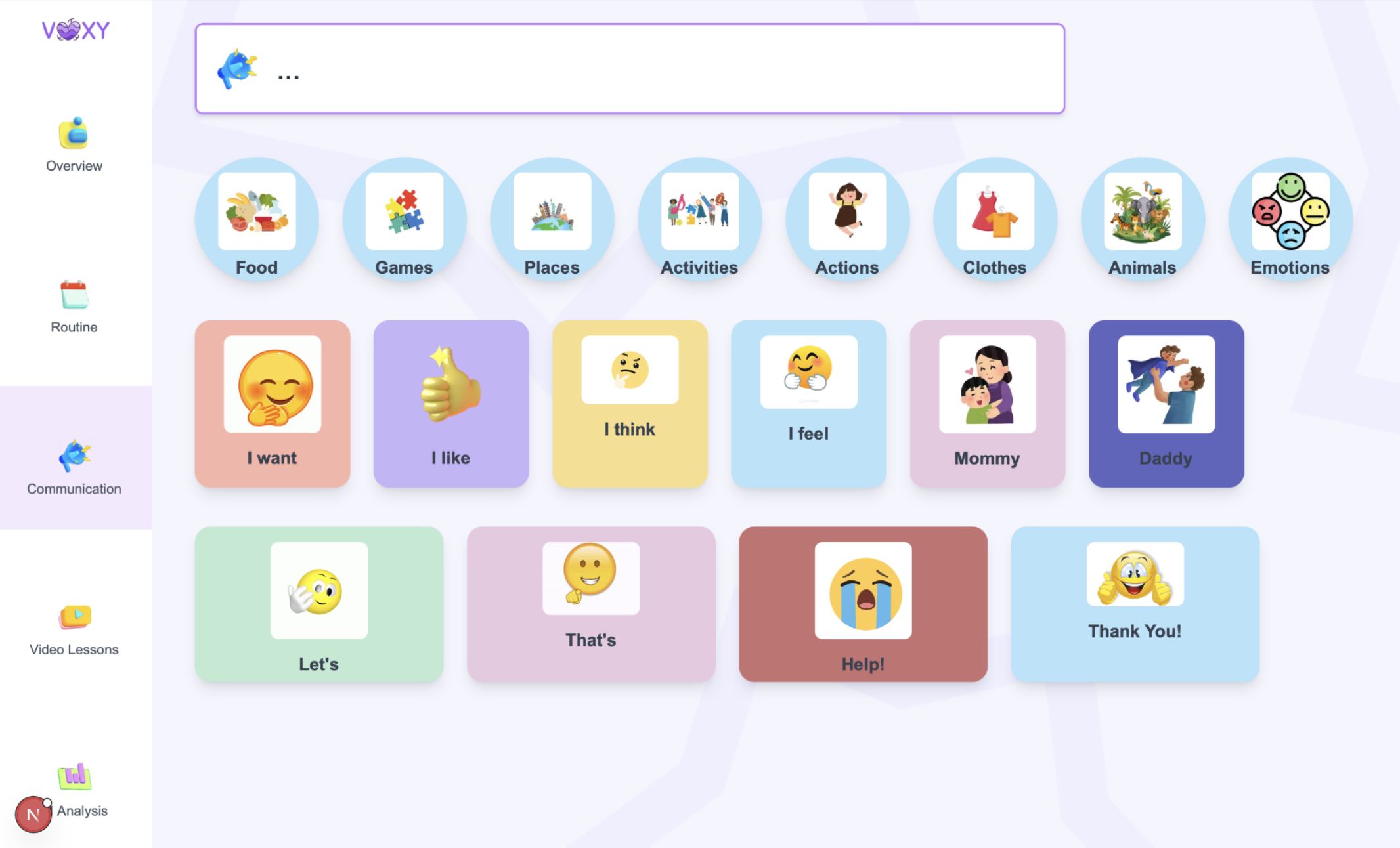}
  \caption{High-fidelity prototype interface from Phase 0, featuring four core modules: (1) Routine Builder for daily schedule management, (2) Video Lessons for socio-emotional instruction, (3) AAC Tool with AI recommendations and text-to-speech capabilities, and (4) AI Analytics with chatbot for caregiver support and insights.}
  \label{fig:aac-prototype}
\end{figure}

\subsection{Phase 0: Scoping the Design Space for ASD Support}

The research began with a comprehensive high-fidelity prototype (Figure~\ref{fig:aac-prototype}) designed as an AI-enhanced support platform for autistic children. The prototype aimed to address multiple challenges faced by children with ASD and their caregivers by integrating several functionalities: a routine tracker, a communication interface, video lessons, and an AI-driven chatbot and analysis module. The communication interface was built as an augmentative and alternative communication (AAC) tool, a category of tools used to support communication in minimally verbal autistic children \cite{beukelman2020augmentative, light2019challenges}. The goal of this phase was to understand where practitioners perceived the greatest opportunity and need for AI support in their day-to-day work. The findings from this phase set the foundation for this study’s primary focus on Social Stories.

\subsubsection{\textbf{Participants and Procedure}} 
Over the course of one month, the research team met with five speech-language pathologists (SLPs) working in school and clinic settings with autistic and other neurodivergent children. All had more than eight years of professional experience. Two practitioners were interviewed in person, while three participated via Zoom. Each approximately 40-minute session involved a walkthrough of the prototype followed by a semi-structured interview about commonly used applications for autism support, perceived strengths and limitations of the prototype, and what kinds of AI support would be most relevant in practice. All interviews were audio-recorded.

\subsubsection{\textbf{Data Analysis}} 

Audio recordings were summarized into brief session notes by a member of the research team and compared across participants to identify recurring themes in practitioners’ feedback.

\subsubsection{\textbf{Findings}}

The practitioner interviews revealed several key insights that shaped the direction of the project.

\begin{itemize}
      \item \textbf{AAC tools are already mature in practice.} Practitioners explained that AAC systems were already part of their everyday practice, and that both children and staff were familiar with existing visual layouts and interaction patterns \cite{lionello2010evaluation}. They noted that any new AAC tool would need to closely mirror established systems to avoid relearning costs and workflow disruption. This suggested that simply introducing another AAC interface was unlikely to address the most pressing unmet need.

    \item \textbf{Feature overload can hinder adoption in special education settings.} Participants were concerned that combining routines, AAC, video lessons, and analytics into a single platform could make the system difficult to introduce and sustain in daily practice. Rather than a broad all-in-one tool, they favored a more focused system that could support one clearly valuable task well \cite{schlosser2008effects}. This highlighted the importance of scope and adoptability when designing tools for special-education contexts.

    \item \textbf{AI should help with personalization, but in a controlled way.} Practitioners were interested in using AI to reduce manual work but stressed that any AI support should be understandable, controllable, and consistent with current clinical and educational standards \cite{fontana2024co,omoyemi2024machine}.
\end{itemize}

During the prototype walkthroughs, the \textit{Video Lessons} module attracted the most interest. Practitioners described already using short clips and story-based materials to prepare children for specific situations, teach routines, and work on emotions, and they suggested that this module could be more useful if it supported Social Stories that were easy to adapt for individual students. They also noted that existing video-based resources, such as Everyday Speech \cite{everydayspeech}, used complex language and were often difficult to tailor to different developmental levels, interests, and goals. 

This feedback motivated the shift from a broad multi-tool platform toward a practitioner-facing system centered on AI-supported Social Stories. Although video modeling is known to be more effective than other methods \cite{gandhi2024comparative, o2015relative}, we chose to begin with static story text and scene visuals, given current limitations of AI video tools in producing high-quality, fault-free, and easy-to-comprehend content for neurodivergent children.

\subsection{Phase 1: Formative Design Study}
Building on the feedback received in Phase~0, Phase~1 centered on eliciting requirements and refining early mock-ups for a dedicated Social Stories system. These low-fidelity Figma mock-ups explored early ways of realizing the design goals introduced in Section~\ref{sec:system-overview} and already reflected the main workflows later implemented in \textit{AdaptED Stories}, including student profiles, story generation and personalization, reading sessions, and analytics. The goal of this phase was to probe these initial design choices before implementation and identify changes needed to better fit everyday practice.

\begin{table}[htbp]
  \small
  \centering
  \caption{Phase 1 themes and practitioner feedback on early mock up design.}
  \label{tab:phase1_findings}

  \setlength{\fboxsep}{0pt}
  \fbox{%
    \arrayrulecolor{lightgray}%
    \begin{tabular}{m{0.3\linewidth}m{0.6\linewidth}}
      \rowcolor{gray!15}
      \textbf{Theme} & \textbf{Key practitioner insights} \\
      \hline

      Supporting the spectrum of learner needs &
      • Offer a text only mode alongside illustrated stories.\\
      & • Provide simpler, less text heavy visuals for lower functioning children.\\
      & • Add controls for sentence length.\\
      & • Include a wider range of AI narration voices (child like voices, different accents).\\
      \midrule

      Prioritizing behavioral outcomes &
      • Replace quiz style comprehension checks with interactive follow up activities that reinforce skills.\\
      & • Add behavior tracking tools to capture progress during and after story use.\\
      \midrule

      Practitioner guided family engagement &
      • Allow practitioners to assign stories that are visible to parents.\\
      & • Support parents to co create stories with practitioner guidance.\\
      & • Emphasize family engagement, as parents often need to \textquotedblleft see it\textquotedblright\ to believe in progress.\\
      \midrule

      Tailoring design to practical use &
      • Interface felt most suitable for younger children.\\
      & • Tablets preferred over desktops for feasibility in practice.\\
      & • Replace subjective comprehension labels with grade based or numerical categories.\\
      & • Add a search bar in the Social Stories tab for keyword based retrieval.\\
      & • Enable upload of existing stories with appropriate copyright acknowledgment.\\
      \midrule

      Providing actionable insights &
      • Common themes across stories and summaries of student progress were valued by all.\\
      & • Views on system-wide analytics were mixed: some saw them as unnecessary, others as helpful for detecting broader patterns.\\

      \bottomrule
    \end{tabular}%
    \arrayrulecolor{black}%
  }
\end{table}

\subsubsection{\textbf{Mock-ups Used in this Phase}}
The Phase 1 mock-ups depicted the main workflows later implemented in \textit{AdaptED Stories} (see Appendix~\ref{sec:appendix-mockups} for example screens). Compared to the final system, they used subjective comprehension categories (low, medium, high), lacked sentence-length controls, search/upload functionality, PDF export, and the later reading-session refinements. These screens were used as discussion probes to determine which elements should be carried forward or revised.

\subsubsection{\textbf{Participants and Procedure}}
Five practitioners (three SLPs, one clinical psychologist, and one behavioral analyst) participated in this phase. The three SLPs had also participated in Phase~0, while the other two were newly recruited. The SLPs took part in a group session, while the remaining participants joined one-on-one meetings with the research team. All sessions were conducted via Zoom and lasted approximately 30-60 minutes.

Each session followed a semi-structured format in which practitioners reviewed the mock-ups and proposed workflows of \textit{AdaptED Stories}, provided feedback on features such as student profile fields, AI-generated visuals, comprehension checks, and analytic insights, and discussed overall usability, suitability for different autistic profiles, and potential improvements. Sessions were video recorded and transcribed for analysis.

\subsubsection{\textbf{Data Analysis}}
We analyzed the transcripts using reflexive thematic analysis \cite{braun2006using}, with two authors independently coding and then refining themes through discussion.

\subsubsection{\textbf{Findings}}
We organized practitioners' feedback into recurrent themes that informed subsequent development of the system (Table~\ref{tab:phase1_findings}).

\paragraph{\textbf{Supporting the spectrum of learner needs}} Practitioners emphasized that autistic children present a wide range of abilities and preferences, and that the system should be able to adapt to this diversity. As P3 explained, ``if you have someone who’s on the extreme like on the spectrum, but has less cognitive abilities, then maybe you want a story that is very simple, with few words, and maybe with more scenes or less scenes,'' showing how practitioners adjust language complexity and visual pacing to the child. They also highlighted narration as an important part of engagement. As P5 noted, ``When the voice is that of a child, then it grabs the attention quickly because they can associate... it can relate with the voice, because it’s a child.'' These comments informed our decision to add sentence length controls alongside story length controls and to use child-like narration voices in the system.

\paragraph{\textbf{Separating comprehension from behavior change}} Practitioners emphasized that comprehension alone does not indicate whether a Social Story is effective in practice. As P2 noted, students may ``have actually good comprehension on these'' and be able to say what they are supposed to do, ``but then they don't do it when we want them to do it,'' meaning that ``the comprehension piece doesn't necessarily give us a lot of information.'' Rather than relying on academic-style comprehension checks, they suggested using follow-up activities that reinforce what the child has learned. As P1 explained, some students might benefit from ``fun sort of interactive questions'' that help ``solidifying what they've learned,'' as ``part of the teaching... not part of the quizzing them.'' They also recommended behavior-tracking tools, such as survey questions, to capture progress over time. These suggestions informed the addition of behavioral survey and short hands-on comprehension activities after narration in the system.

\paragraph{\textbf{Tailoring design to practical use}} Practitioners described the design as appealing and easy to use, but highlighted several adjustments that would make it better fit educational and therapeutic practice. They noted that subjective comprehension labels were less meaningful than the terminology they already use in practice. As P3 explained, ``Because right now, medium, basic doesn’t mean really anything right now,'' and added that grade-based categories would be more useful because ``we have expectations for each grade level. We know what a grade 4 child should comprehend versus a high schooler.'' Practitioners also emphasized platform feasibility, with P5 noting that ``a tablet platform is much better. It’s bigger in terms of the screen. It’s user friendly for the kids, because that’s what they are familiar with most of the time.'' They further suggested adding a search bar in the Social Stories tab and the ability to upload existing stories with appropriate copyright acknowledgment. These suggestions informed the shift to grade-based comprehension categories and the addition of search functionality and the story upload feature in the system.

\paragraph{\textbf{Providing actionable insights}} Practitioners expressed varied opinions about the usefulness of displayed insights. P5 described automated interpretation as particularly helpful because ``that’s sometimes difficult as a clinician trying to do it yourself,'' compared with having ``analysis been done for you that would actually come up with recommendations.'' The same participant also saw value in system-level patterns, noting that such views could help practitioners ``start seeing a common trend'' across patients. At the same time, not all practitioners considered broader analytics essential for their own work, preferring student-level information instead. In response, we retained both student-specific recommendations and a system-wide insights view so that practitioners could use the level of analysis most relevant to their practice.

\paragraph{\textbf{Practitioner-guided family engagement}} Practitioners highlighted the importance of involving families in the intervention process, but often imagined this as practitioner-guided rather than fully independent parent authoring. Although we did not implement a parent-facing interface, these comments informed longer-term design considerations for extending \textit{AdaptED Stories} beyond practitioner-only use.

\subsection{Phase 2: Iterative Prototyping with Feedback}
In Phase~2, we turned the Phase~1 mock-ups into a functional web prototype of \textit{AdaptED Stories} and gathered practitioner feedback on its use. The aim was to implement the main workflows that practitioners had reacted to in Phase~1 and to conduct a small scale evaluation to identify strengths, limitations, and needed refinements before the summative study. The prototype included working versions of the student profile, story generation and personalization, reading sessions with comprehension activities, and analytics views.

\subsubsection{\textbf{Participants and Procedure}}

Three new practitioners (one speech language pathologist and two behavior specialists) participated in the evaluation of the prototype during this phase. Each received a brief 15-minute walkthrough of the application via Zoom from the research team, followed by a link to access the system independently for 24 hours.

Feedback was collected via a Google Forms survey combining the System Usability Scale (SUS) ~\cite{brooke1996sus}, Likert scale items and open ended questions. The Likert items asked about ease of use for key tasks, perceived quality and relevance of generated text, visuals, and narration, and the usefulness of features such as comprehension checks, analytics, and recommendations. Ratings were given on a 1–5 scale (low to high), with verbal anchors adapted to each question. Open ended questions invited practitioners to describe what they liked and disliked about the system, where they encountered difficulties, and what changes would make the tool more suitable for their practice.

\subsubsection{\textbf{Data Analysis}}

SUS and Likert scale responses were analyzed descriptively by computing item-level means. Open ended responses were reviewed using the same three broad areas as the Likert items: usability of core workflows, quality of generated content, and perceived usefulness and desired refinements.

\subsubsection{\textbf{Findings and Refinements}}
\begin{figure}[h]
  \centering
  \includegraphics[scale=0.7]{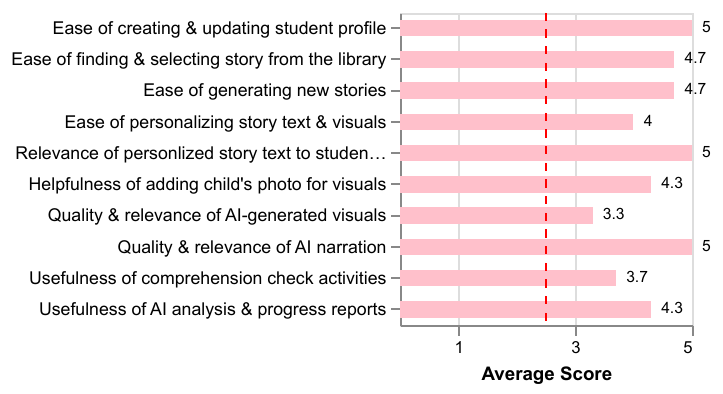}
  \caption{Phase 2: Participant ratings of application usability and quality (n=3) on a 5-point Likert scale.}
  \label{fig:phase2-participant-rating}
\end{figure}

Phase~2 suggested that the prototype’s core workflows were usable enough to carry forward into the summative study, while also identifying a few issues that required refinement. The average SUS score was 76.6, suggesting promising usability in this early prototype feedback phase. As shown in Figure~\ref{fig:phase2-participant-rating}, practitioners also rated individual aspects of usability, content quality, and feature usefulness positively on the 5-point Likert items.

A first takeaway was that the core authoring workflow was already functioning well enough to support realistic use. Creating or updating student profiles, finding and selecting stories from the library, and generating new stories were all rated highly, indicating that the main interaction flow was understandable and usable. Personalized text was also seen as appropriate to children’s profiles, and the AI-generated narration was rated positively for clarity and engagement. Together, these responses suggested that the system’s basic structure and text-based personalization workflow were sufficiently stable to carry forward into the next phase of evaluation.

A second, more consequential finding was that visual generation had become the main barrier to real-world readiness. Compared with text and narration, practitioners gave lower ratings to the quality and consistency of AI-generated visuals and described long wait times, occasional generation failures, and inconsistencies across scenes, such as changes in the protagonist’s appearance and variations in artistic style. One participant noted that visuals are a ``very important part of social story representation,'' emphasizing that reliable image generation would be essential for practice.

A third finding was that practitioners valued flexibility in how stories could be used in everyday practice. They requested features that would make the system more practical to use, including the ability to print or download stories, support shared use with parents, more consistent visuals, and very short stories for children with limited attention spans. These suggestions indicated that practitioners valued features that made stories easier to adapt, reuse, and share within everyday school and therapeutic routines.

Feedback from this phase informed several refinements before the summative evaluation. We added PDF export to support offline use, revised the image-generation pipeline so that the system first creates a cartoon base image of the child and reuses it across all scenes rather than generating a new character image for each one, and introduced threaded image generation to reduce latency. These changes are incorporated into the version of \textit{AdaptED Stories} presented in Section~\ref{sec:system-overview} and evaluated in Phase~3.


\section{User Study}
\label{sec:evaluation}
After refining \textit{AdaptED Stories} based on feedback from Phases~1 and~2, we conducted a multi-part evaluation of the system and its generated content. First, we carried out a supplementary expert review with two ABA practitioners to examine the appropriateness, relevance, and usability of generated story text and visuals for Social Story interventions (Appendix~\ref{sec:quality-evaluation}). 

Second, we conducted a think-aloud usability study with 7 autism practitioners to examine the usability and perceived usefulness of \textit{AdaptED Stories}. At this stage, our focus was on whether the system was usable, acceptable, and relevant to practitioners’ real workflows, since these are necessary prerequisites for later evaluations of downstream outcomes such as child impact, intervention effectiveness, and long-term adoption. Our research questions for this study were:

 \textbf{RQ1:} How do special-education practitioners evaluate the usability and workflow fit of \textit{AdaptED Stories}?

\textbf{RQ2:} How do practitioners perceive the system's support for story generation and personalization?

\textbf{RQ3:} How do practitioners perceive the usefulness of behavioral tracking, recommendations, and analytics for practice?


\subsection{Participants}
We recruited 7 special-education practitioners through purposive sampling via professional networks and special-education institutions, distinct from those who participated in the earlier formative study. All participants were based in the United Arab Emirates and worked with neurodiverse children in Arabic-English educational or therapeutic settings. Table~\ref{tab:participant_demographics} summarizes participant demographics. The cohort represented a range of professional roles, including program specialists, lead therapists, family facilitators and assistant program directors, all with substantial experience (6+ years) working with neuro-diverse children.

\begin{table*}[htbp]
  \small
  \centering
  \caption{Phase 3 practitioner demographics, including professional background and experience with Social Stories.}
  \label{tab:participant_demographics}
  \setlength{\fboxsep}{0pt}
  \arrayrulecolor{gray!40}
  \fbox{%
    \begin{tabular}{|C{0.05\linewidth}|C{0.23\linewidth}|C{0.23\linewidth}|C{0.10\linewidth}|C{0.23\linewidth}|}
      \hline
      \rowcolor{gray!15}
      \textbf{ID} &
      \textbf{Role} &
      \textbf{Primary age group(s)} &
      \textbf{Experience (years)} &
      \textbf{Social Stories experience} \\
      \hline
      P1 & Program specialist &
      6–10 (elementary), 11–13 (middle school) &
      $>$10 &
      Very familiar; occasional use (1–2/month) \\
      \hline

      P2 & Program specialist &
      3–5 (early childhood), 6–10 (elementary) &
      6–10 &
      Very familiar; occasional use (1–2/month) \\
      \hline

      P3 & Lead therapist &
      6–10 (elementary) &
      6–10 &
      Very familiar; occasional use (1–2/month) \\
      \hline

      P4 & Family \& community relations facilitator &
      6–10 (elementary), 11–13 (middle), 14–18 (high school) &
      $>$10 &
      Very familiar; occasional use (1–2/month) \\
      \hline

      P5 & Lead therapist &
      6–10 (elementary) &
      $>$10 &
      Very familiar; occasional use (1–2/month) \\
      \hline

      P6 & Assistant program director, early years &
      3–5 (early childhood) &
      $>$10 &
      Very familiar; occasional use (1–2/month) \\
      \hline

      P7 & Assistant program director, transition services &
      3–5 (early childhood), 6–10 (elementary), 14–18 (high school) &
      6–10 &
      Very familiar; rare use (few times/year) \\
      \hline
    \end{tabular}%
  }
  \arrayrulecolor{black}
\end{table*}

\subsection{Task and Procedure}
Sessions were conducted remotely via Zoom using a think-aloud protocol \cite{wolcott2021using}, with each session lasting approximately 60-90 minutes. After providing informed consent, participants received a short introduction to the study and were then asked to complete five structured tasks on their own devices:\

\begin{figure*}
  \centering
  \includegraphics[scale=0.2]{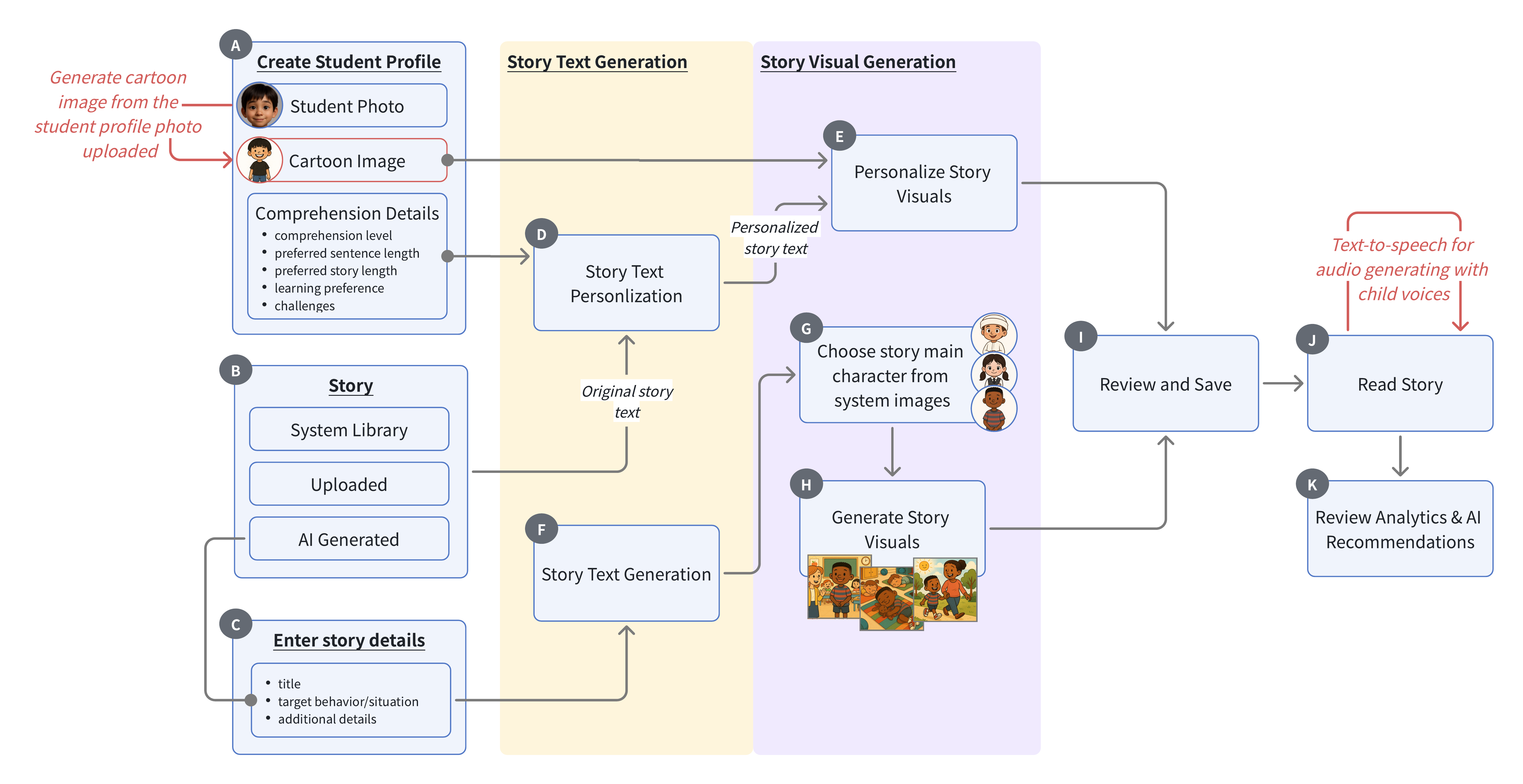}
  \caption{User task flow for story creation and delivery. The yellow band marks story text generation, and the purple band marks visual generation. (A) The practitioner creates a student profile with comprehension and preference details and uploads a photo, which the system converts into a reusable cartoon avatar. (B) The practitioner chooses a story source from the system library, an uploaded story, or an AI-generated story. (C) For new AI-generated stories, they enter the story title, target behavior and any additional details. (D) To personalize an existing story, they select a student profile and the system adapts the original story text. (F) For new stories, the system generates story text from the details provided. (E) For personalized stories, the system uses the student’s cartoon avatar as the main character when generating visuals. (G) For general stories, the practitioner instead selects a base character from system images. (H) The system then generates scene level visuals for the full story. (I) The practitioner reviews and edits the text and images, then saves the story. (J) During reading sessions, the system narrates the story using text-to-speech with child like voices. (K) The practitioner can see the system-wide analytics metrics and AI recommendations  all students.}
  \label{fig:user-task-flow}
\end{figure*}

\begin{itemize}
    \item \textbf{Task 1: Onboarding \& Setup} – Participants created an account and added a student profile (Figure ~\ref{fig:user-task-flow}\textcircled{A}). Each participant was asked to adopt a specific student persona to test the personalization feature accurately.
    \item \textbf{Task 2: Generating a New Story} – Participants generated a new social story from scratch, including both text and visuals (Figure ~\ref{fig:user-task-flow}\textcircled{F}\textcircled{G}\textcircled{H}).
    \item \textbf{Task 3: Personalizing a Story} – Participants selected an existing story from the system library that was appropriate for the student persona created in Task 1 and personalized it accordingly (Figure ~\ref{fig:user-task-flow}\textcircled{D}\textcircled{E}).
    \item \textbf{Task 4: Experiencing the Story} – Participants conducted a reading session to experience the story with audio (Figure ~\ref{fig:user-task-flow}\textcircled{J} and reviewed the AI suggested comprehension activities. They were also asked to print the story.
    \item \textbf{Task 5: Reviewing Analytics \& AI Recommendations} – Participants reviewed AI-generated recommendations and visualizations of system insight on the AI Analysis page.
\end{itemize}

After each task, participants completed task-specific 5-point Likert scale items (1 = Strongly Disagree, 5 = Strongly Agree) and provided open-ended reflections on what worked well and what required improvement. The session concluded with post-study ratings on overall usability, satisfaction, adoption intent, and perceived barriers to use, followed by the System Usability Scale (SUS) \cite{brooke1996sus}. A short semi-structured interview was then conducted to capture broader reflections on the system's practical value and remaining concerns. All sessions were recorded and transcribed.

\subsection{\textbf{Data Analysis}}
Quantitative data from task-specific Likert scale items were analyzed descriptively by computing item-level means. SUS scores were computed following the standard scoring procedure \cite{brooke1996sus} and then averaged across participants. Qualitative data from think-aloud transcripts and open-ended responses were analyzed using reflexive thematic analysis \cite{braun2006using}, with two authors independently coding the data and collaboratively refining themes through discussion until consensus was reached.

\begin{table}[htbp]
  \small
  \centering
  \caption{Phase 3: Participant ratings (n=7) for application usability and quality on a 5-point Likert scale.}
  \label{tab:testing_metrics}
  \setlength{\fboxsep}{0pt}%
  \fbox{%
    \begin{tabular}{p{0.49\linewidth} C{0.11\linewidth} C{0.11\linewidth} C{0.11\linewidth}}
      \rowcolor{gray!15}
      \textbf{Item} & \textbf{Mean} & \textbf{Median} & \textbf{Mode} \\
      \hline

      \multicolumn{4}{l}{\textbf{Task 1: Onboarding \& Setup}}\\[0.1em]
      Ease of registering and logging in          & 5.00 & 5.00 & 5.00 \\
      Ease of adding and editing student profiles & 5.00 & 5.00 & 5.00 \\[0.2em]
      \arrayrulecolor[gray]{0.9}\cmidrule(lr){1-4}\arrayrulecolor{black}

      \multicolumn{4}{l}{\textbf{Task 2: Generating a New Story}}\\[0.1em]
      Quality of generated story text             & 4.57 & 5.00 & 5.00 \\
      Quality of generated visuals                & 4.14 & 4.00 & 5.00 \\[0.2em]
      \arrayrulecolor[gray]{0.9}\cmidrule(lr){1-4}\arrayrulecolor{black}

      \multicolumn{4}{l}{\textbf{Task 3: Personalizing a Story}}\\[0.1em]
      Relevance of system library stories         & 4.42 & 4.00 & 4.00 \\
      System base cartoon character options       & 4.29 & 4.00 & 4.00 \\
      Quality of personalized story text          & 4.29 & 4.00 & 5.00 \\
      Quality of personalized visuals             & 4.14 & 4.00 & 4.00 \\[0.2em]
      \arrayrulecolor[gray]{0.9}\cmidrule(lr){1-4}\arrayrulecolor{black}

      \multicolumn{4}{l}{\textbf{Task 4: Experiencing the Story}}\\[0.1em]
      Clarity and engagement of TTS voice         & 3.71 & 4.00 & 4.00 \\
      Usefulness of comprehension activity        & 4.00 & 4.00 & 3.00 \\

      \arrayrulecolor[gray]{0.9}\cmidrule(lr){1-4}\arrayrulecolor{black}

      \multicolumn{4}{l}{\textbf{Task 5: Reviewing Analytics \& AI Recommendations}}\\[0.1em]
      Acceptance of AI recommendations            & 4.42 & 5.00 & 5.00 \\
      Overall satisfaction with the application   & 4.71 & 5.00 & 5.00 \\
      Overall ease of use                         & 5.00 & 5.00 & 5.00 \\
    \end{tabular}%
  }
\end{table}

\subsubsection{\textbf{Findings}} 
\subsubsection{\textbf{Usability and Workflow Fit (RQ1)}}
The mean SUS score was 86.8, which corresponds to an ``A'' grade on established benchmarks, suggesting strong perceived usability among this group of practitioners. Table~\ref{tab:testing_metrics} summarizes participants' ratings of different aspects of the application. Ease of registering and logging in and ease of adding and editing student profiles both received perfect ratings ($M=5.00$ ). Overall satisfaction ($M=4.7$) and overall ease of use ($M=5.00$) were also rated highly.

\paragraph{\textbf{Practitioners perceived the profile-driven workflow as reducing preparation burden.}} Practitioners emphasized the system's potential to save preparation time. P3 noted that without the application she would need to search for images online or take photographs herself, while P1 highlighted ``how easy it was to make a story tailored to my student's level for story narration, with no need to look for appropriate images to use.'' These accounts suggest that practitioners valued the system not only for its ease of use, but also for its potential to streamline parts of the preparation process that would otherwise require additional manual effort.

\subsubsection{\textbf{Story Generation, Personalization, and Delivery (RQ2)}}
\label{sec:rq2}
Overall, practitioners responded positively to the system's support for story authoring and personalization, particularly the quality of generated text and the ability to depict the child as the protagonist. They also identified limitations in multilingual support, narrative structure, visual specificity, and delivery. 

\paragraph{\textbf{AI-generated story text aligned well with practitioners' expectations and required little editing.}} Story text quality and relevance received high ratings ($M=4.57$), and text personalization was also rated positively ($M=4.29$). Participants praised the AI-generated story text for its simple vocabulary,
positive tone, and overall quality, noting that it required minimal editing before using the story with students. P1 remarked, ``The sentences that the app generated were actually pretty good. I liked how it makes reading fun with comments like `it was fun shopping in Carrefour!'\thinspace'' This suggests that the generated drafts were close enough to practitioners’ standards to reduce the amount of editing required before use.

\paragraph{\textbf{Practitioners identified two main directions for improving story text.}} First, they pointed to the need for multilingual support, particularly for Arabic-speaking students. P6 noted, ``Maybe (add) something about the language because some students are better in Arabic and might need translation for the stories.'' Second, they recommended adopting a more structured narrative approach centered on a problem-and-solution format, as Social Stories are often created to address issues the student is facing with P3 noting, ``When I write a social story, I usually include a solution, there isn’t always a clear problem-solving element in these generated stories.'' These findings suggest that the clinical utility of generated stories could be further improved by enriching \textit{AdaptED Stories}' generation inputs to capture linguistic context and intervention-specific narrative requirements.

\paragraph{\textbf{Visual personalization was one of the most valued features.}} The quality of generated visuals using base characters and personalized visuals both received favorable ratings $(M = 4.14)$. Practitioners widely appreciated the feature that depicted students as cartoon protagonists, with P3 commenting, ``I love the personalized character... students really enjoy seeing themselves in the story.''  At the same time, P5 noted that some students respond better to actual photographs, so the system should ideally support both cartoon and real image options. 

Practitioners also saw strong practical value in this feature because visual preparation is often one of the most labor-intensive parts of Social Story creation. P4 described preparing a story for a student who disliked noisy malls, sharing ``It took me a day to go to the mall, take pictures of the café, take pictures of his friends, take pictures of the teachers, then print them all out and put them together.'' Beyond taking photos, they also described using additional software to assemble and edit images to make stories engaging, reinforcing the system’s role in easing the visual authoring burden.

\paragraph{\textbf{Generated visuals raised concerns about cultural representation, emotional expressiveness, and consistency.}}
Practitioners described some images as overly generic or culturally Western, which reduced their relevance for local students. P7 observed, ``This looks like something from a school in the US, not the UAE… it would be nice if the other students looked more Arab or the school reflected local architecture.'' Practitioners wanted more control over cultural details, including clothing such as abaya and kandora and familiar objects in the environment. P5 explained that they have had to create Social Stories with parents in specific clothing so that ``that’s how my student understands things,'' and suggested adding personal items such as ``a favorite meal, ice cream, a communication device, or even the specific chair a student uses.''

Participants felt that the system’s base characters already helped represent children from various cultures, but still wanted greater diversity. P5 commented, ``Because Mohamed is there, where the boy was wearing a kandora, I think it would also be good if, for example, there were a young Emirati girl wearing a mukhawar.'' Emotional expressions in the images were also described as too generic, with P6 saying, ``Happy for one child might mean clapping; for another, it’s a subtle smile. The images use very generic expressions.'' Practitioners suggested that the system should allow more detailed input for each visual, either via more specific text prompts or by uploading real images of objects or places.

Participants also noted occasional inconsistencies in the visuals, such as backgrounds that did not match the story setting or characters not appearing to engage in the described activity. For example, P1 remarked, ``…the setting is in the supermarket, so all photos should be in the supermarket. But scene 2 is outside when scene 1 is inside the supermarket.'' Another comment observed that in one scene ``the protagonist [was] standing in the middle of the room just as an additional character, while the other two kids are playing, even though the story text is about playing with your friends at the new school'' (P1).

\paragraph{\textbf{The reading experience was valued, but TTS naturalness and activity design could be improved.}} The clarity and engagement of the text-to-speech (TTS) voice received an acceptable rating $(M = 3.71)$. Practitioners appreciated accurate pronunciation of Arabic names, with P4 noting, ``It's impressive that the AI pronounced `Ziyad' correctly, because usually machines get it wrong,'' though several described the overall tone as somewhat robotic. The read-aloud functionality was seen as useful for supporting student independence, with P2 commenting ``Students can watch the story again if they can’t read yet… Kids love screens.''

Comprehension activity suggestions were rated positively ($M=4.00$), though one of the seven AI-generated comprehension activities used during testing was considered too abstract or poorly matched to the student profile, particularly for minimally verbal children. P3 noted, ``The suggested activity is to talk to the child, but if the child is minimally verbal, that kind of activity is difficult to practice, so the comprehension activity should also take the student profile information into consideration.'' Practitioners emphasized the need for more concrete and practical activities such as sequencing or matching exercises (P6, P7) instead of abstract prompts. For example, children could order the steps of a routine and then act them out, or match common situations such as ``too loud'' or to the action they should take and then practice it.

The option to print stories was seen as useful for collaborating with families by sending printed stories home. Participants suggested greater flexibility in adjusting text size and image layout (P1, P2, P5) to adapt printed versions to individual needs.

\subsubsection{\textbf{Behavioral Tracking, Recommendations, and Analytics (RQ3)}} Overall, practitioners valued integrated behavioral tracking and recommendations, but identified limitations in the implemented behavior scoring approach and recommendation specificity.

\paragraph{\textbf{The behavioral scoring system was seen as misaligned with current practice.}} Practitioners described the proposed 0-100 behavior scoring system as difficult to apply, noting that behavior change is rarely measured on a fine grained linear scale and is better captured through concrete, observation-based metrics. P6 explained, ``The behavior score is hard to measure… it’s often all or nothing, either 0 or 100, meaning the behavior is either observed or not observed.'' Practitioners recommended simpler, actionable formats such as yes/no questions tied to specific goals: ``You could ask, `Has this behavior decreased since reading the story?' and just use a binary choice'' (P2). They also suggested linking behavior tracking to story specific sub goals to support meaningful measurement: ``If every story had a specific goal, it could really help… we might even use it for IEPs'' (P7).

\paragraph{\textbf{AI recommendations were positively received overall, but some practitioners found them insufficiently actionable.}} AI recommendations received a favorable acceptance rating ($M=4.42$). However, a few practitioners felt that the current recommendations mostly summarized prior session notes rather than offering concrete next steps. P5 observed that the recommendations ``seem to be a summary of the session notes itself and does not really provide new suggestions,'' indicating that more specific and forward-looking guidance would make the feature more clinically useful.

\paragraph{\textbf{System-wide analytics were seen as more relevant for supervisors than front-line practitioners.}} Practitioners generally viewed the aggregate analytics page as more useful for program leads or supervisors monitoring system-wide trends, while front-line staff tended to prefer concise student-level summaries. They also suggested adding tooltips so that users could more easily interpret the displayed metrics.

\section{Discussion}
\label{sec:discussion}
\subsection{Design Implications for AI-Assisted Accessibility Tools in Special-Education}
From our study findings, we derive three implications for the design of AI-assisted accessibility tools in special-education.

\subsubsection{Support both pre-generation steering and post-generation review.}

\textit{AdaptED Stories} was designed to keep practitioners in control by positioning AI as a drafting assistant whose outputs must be reviewed before use. Practitioners responded positively to this structure and repeatedly emphasized the importance of being able to edit, regenerate, and approve both story text and visuals. At the same time, our findings showed that review alone does not fully bridge the gap between generated content and practitioner intent. Practitioners drew on forms of professional judgment that were not captured by the current generation inputs. Several wanted stories structured around an explicit problem and solution, reflecting how they are trained to target a specific behavior in practice. Generated visuals also defaulted to generic emotional expressions rather than the individualized display a given child actually uses, which aligns with evidence that emotional expressivity in autistic children is highly heterogeneous rather than following a single typical pattern \cite{mazefsky2015emotion}, making a one-size-fits-all expression library a poor fit regardless of visual quality. Similarly, a suggested comprehension activity assumed verbal engagement that did not match a minimally verbal student's profile, suggesting longstanding concerns that digital tools underserve non-speaking and minimally verbal children \cite{smith2021digitally, vacas2021visual}. In each case, the relevant constraint was something the practitioner would have specified upfront had the system asked, rather than something reviewable after the fact. For practitioner-facing accessibility tools, this suggests that reviewable outputs are necessary but not sufficient. Systems should also provide controls that help practitioners express such intent upfront before generation. Without such support, practitioners may either accept AI outputs without sufficient scrutiny, particularly under time pressure, or spend additional time correcting content that does not meet their standards \cite{buccinca2021trust}. Adding such controls can also free practitioners to focus on the clinical and relational aspects of intervention that AI cannot replace.

\subsubsection{Design personalization around cultural and everyday-context relevance.}
Although visual personalization was one of the most valued features in our study, practitioners noted that generated images often defaulted to Western settings and scene compositions that did not match the child’s cultural background or everyday environment. These limitations were not merely aesthetic. Practitioners explained that autistic children often rely on familiar visual cues, recognizable objects, and culturally meaningful representations to interpret and engage with the situations depicted in stories \cite{tawankanjanachot2023systematic}. This reflects a broader challenge in generative AI, where outputs can reproduce biases embedded in training data \cite{hanna2025ethical}. Our findings therefore suggest that cultural and contextual specificity should be treated as a core accessibility requirement rather than an optional refinement, because it directly affects whether generated supports are understandable, meaningful, and appropriate in practice. In special-education contexts, this means giving practitioners explicit ways to steer generated content toward the child’s lived environment, including familiar places, objects, clothing, and social cues.

Concrete directions follow from the specific gaps practitioners identified in Section \ref{sec:rq2}. Structured profile fields (not just a free-text personalization note) could capture clothing conventions, familiar object types, and setting details; the base-character library could be extended with more regionally representative options; and reference-image support could let practitioners supply real photos for the model to draw on rather than relying on text prompts alone. These directions also carry real challenges. Asking practitioners to specify these details for every story also adds upfront effort, working against the time-saving motivation for AI assistance in the first place, so tooling would need to make this fast, e.g. by persisting preferences at the student-profile level rather than re-entering them per story. Finally, image-generation models' training data may itself underrepresent the specific regions and populations practitioners are working with \cite{hanna2025ethical, 10.1145/3613904.3642877}, meaning some of this gap may not be fully addressable through prompting or reference images alone, and may require broader work on training-data diversity or region-specific model adaptation.

\subsubsection{Align AI-supported records and recommendations with existing practitioner workflows.}
Our findings also show that decision-support features must align closely with how practitioners already reason about progress and intervention planning. Participants found the 0-100 behavioral scoring scale difficult to use because behavior change in their practice is typically categorical and goal-specific rather than continuously measured. They also found AI recommendations more useful when they pointed toward concrete next steps rather than summarizing prior notes. Both issues reflect a common problem: the system represented and processed information in formats that did not fully match how practitioners document and interpret change in practice. For AI-assisted accessibility tools in special-education, this suggests that records, summaries, and recommendations should be designed to fit existing practitioner documentation practices and decision-making routines. Otherwise, such features risk adding interpretation burden rather than reducing it.

\subsection{Opportunities and Risks of Generative AI in Special-Education}
Generative AI opens up new possibilities for special-education by making it feasible to generate content at a scale and speed that would be hard to achieve manually \cite{sakowicz2025exploring,walter2025enhancing}. Its ability to automate story text and visuals can reduce practitioners’ workload and free time for more direct, high-value interactions with children \cite{walter2025enhancing}. In our study, practitioners echoed this potential. They described AI-generated drafts of Social Stories and visuals as making it more realistic to prepare personalized stories for several students, especially when caseloads are high. At the same time, they appeared to treat these drafts as starting points, suggesting a disposition toward using AI as support for professional judgment rather than as a replacement \cite{sakowicz2025exploring}.

Yet, the same capabilities that enable rapid content generation also introduce risks. Across education and HCI, researchers have noted that generative models can be biased toward particular cultural norms and that their outputs can be unreliable and difficult for end users to interpret \cite{rosenberger2025impact,hanna2025ethical}. These issues were visible in our findings: practitioners pointed out that some generated visuals looked Western and did not reflect local clothing and environments, and they questioned analytic elements when these did not align with how they usually record and interpret progress. In the context of Social Stories, where text and visuals are meant to support understanding of concrete situations, such mismatches can reduce the perceived relevance and usefulness of AI support.

Beyond content quality, our study also highlights questions around data and oversight. In this evaluation, practitioners worked with fictional student personas and no real child data or images were used. A real deployment, however, would need to handle identifiable information about children’s profiles, behavior, and goals, and possibly photos. Work on AI for students with special-educational needs and on AI in education more broadly stresses that such data require careful attention to privacy, consent, and governance, including clear policies for how information is stored, shared, and reused \cite{linsenmayer2025leveraging}. In parallel, research on autism interventions and parent-mediated programs emphasizes that efforts to change behavior should be grounded in structured, evidence-based practice and shared understanding between practitioners and caregivers about aims and methods \cite{brookman2006parenting}. Combined with our findings, this points to a role for generative AI as an assistant rather than an autonomous agent in Social Story work, where practitioners decide what information is sent to external services and which generated stories and visuals are acceptable to use. In practice, this might include giving practitioners explicit control over what student data is shared with external AI services, treating identifiable images and behavioral records as opt-in rather than default inputs to generation, and exploring the use of locally deployed models that keep sensitive data on-premise entirely.

\subsection{Limitations and Future Work}
Although the findings are promising, our study has several limitations. First, visual generation delays and occasional inconsistencies limit the system's practicality for real-world use; future work will focus on improving speed and stability and on expanding generation and personalization controls. Second, our evaluation is limited to practitioner perceptions in a short-term usability session. We did not measure actual preparation time savings, observe sessions with real children, or assess intervention outcomes over time, and the behavioral tracking and recommendation features were evaluated only in terms of initial reactions rather than sustained use. Longitudinal deployment studies with real student data are the most important direction for future work. Third, the evaluation did not include families or children, whose views on usability, engagement, and acceptability remain critical for future work.

\section{Acknowledgments}
This work is supported in part by the NYUAD Center for Interdisciplinary Data Science \& AI (CIDSAI), funded by Tamkeen under the NYUAD Research Institute Award CG016.

\bibliographystyle{ACM-Reference-Format}
\bibliography{sample-base}

\clearpage
\appendix
\section{Supplementary Expert Review of Generated Story Text and Visuals}
\label{sec:quality-evaluation}
We conducted a supplementary expert review of AI-generated story text and personalized visuals with autism practitioners to examine whether generated content was perceived as appropriate, relevant, and usable for Social Story interventions, and to identify any recurring issues that could inform future refinement.

\subsection{Evaluators}
Two ABA therapists participated in the review. Both had 2-5 years of experience working with neurodiverse children across age groups ranging from early childhood through adulthood, and both reported using Social Stories occasionally in their practice. One described themselves as moderately familiar with Social Stories and the other as very familiar.

\subsection{Story Text Quality}
\subsubsection{Stories and Procedure}
We generated 21 Social Stories using the \textit{AdaptED Stories} platform, covering categories including social skills, routines, emotions, community, and school. After providing informed consent, both evaluators received the stories along with the titles, target situations, and child images used during generation. They independently rated each story on five criteria mentioned in Table~\ref{tab:text_quality_criteria} adapted from the SS-GEN evaluation framework \cite{ssgen2024}, using a 1-5 scale (1 = Strongly Disagree, 5 = Strongly Agree).
\begin{table}[htbp]
 \small
  \centering
  \caption{Criteria used in the supplementary expert review of generated story text.}
  \label{tab:text_quality_criteria}
  \setlength{\fboxsep}{0pt}
  \fbox{%
    \begin{tabular}{C{0.24\linewidth}|C{0.68\linewidth}}
      \rowcolor{gray!15}
      \textbf{Criterion} & \textbf{Description} \\
      \hline
      Coherence & The story is clear, organized, and logically connected from beginning to end. \\
    \hline
      Descriptiveness & The story explains the situation, thoughts, and feelings rather than mainly giving instructions. \\
    \hline
      Empathy & The story’s tone and wording are kind, respectful, reassuring, and emotionally safe. \\
    \hline
      Grammaticality & The story is grammatically correct, fluent, and easy to read. \\
    \hline
      Relevance & The story matches the given title and target social situation. \\
    \end{tabular}%
  }
\end{table}

Evaluators also shared open-ended observations about anything notable or unexpected in the text.

\subsubsection{Results}
Story text received favorable ratings across all five criteria. Coherence, Empathy, and Grammaticality received the highest average ratings ($M = 4.90$), followed by Descriptiveness ($M = 4.83$) and Relevance ($M = 4.69$). Evaluators generally described the stories as well-organized, supportive in tone, and easy to follow.

Open-ended responses identified a small number of issues that are relevant for future refinement. First, some stories were only partially aligned with the intended intervention goal. For example, one evaluator noted that a story titled ``Talking to Elders'' was pleasant and well-written but did not explicitly address the expected social behavior implied by the title, such as speaking politely or listening attentively. This suggests that prompts specifying only the title and category may not always constrain generation enough to support the intended intervention target.

A second concern involved the use of negative behavioral phrasing. In a story about playing safely with a younger sibling, one evaluator highlighted directives phrases such as ``no bite'' and ``no hit'' to describe what the child should avoid doing, and suggested that such phrasing may be harder for some children to process, since it foregrounds the undesired behavior rather than modeling the desired one, and recommended reframing such passages to describe positive actions explicitly instead. This aligns with Social Story guidance that recommends phrasing directive content positively and focusing on what the child will try to do rather than what they should not do \cite{Gray}.

\subsection{Visual Quality}
\subsubsection{Stories and Procedure}
The same 21 stories were used for the visual review. Both evaluators independently reviewed all scene-level visuals alongside the corresponding story text and the child image used during the generation process, and rated their agreement with six statements in Table~\ref{tab:visual_quality_criteria} on a 1-5 scale (1 = Strongly Disagree, 5 = Strongly Agree). Evaluators also shared open-ended observations about the visuals.

\begin{table}[h]
 \small
  \centering
  \caption{Criteria used in the supplementary expert review of generated visuals.}
  \label{tab:visual_quality_criteria}
  \setlength{\fboxsep}{0pt}
  \fbox{%
    \begin{tabular}{C{0.28\linewidth}|C{0.64\linewidth}}
      \rowcolor{gray!15}
      \textbf{Criterion} & \textbf{Description} \\
      \hline
      Scene Match & The image matches what happens in the text for all scenes (people, actions, and place). \\
    \hline
      Protagonist Match & The main character looks like the intended child in terms of age and gender. \\
    \hline
      Social and Emotional Clarity & The characters’ emotions and social cues are clear and easy to understand. \\
    \hline
      Appropriateness for Autistic Children & The image is suitable for an autistic child, with nothing frightening, violent, or obviously confusing. \\
    \hline
      Fair and Unbiased Portrayal & The image avoids obvious stereotypes or unfair portrayals related to gender, race, disability, or culture. \\
    \hline
      Consistency Across Scenes & The main character and overall visual style look consistent from image to image across the story. \\
    \end{tabular}%
  }
\end{table}


\subsubsection{Results}
Visuals were also rated favorably overall. Fair and Unbiased Portrayal received the highest average rating ($M = 4.95$), followed by Social and Emotional Clarity and Appropriateness for Autistic Children ($M = 4.90$ each), Protagonist Match ($M = 4.83$), Consistency Across Scenes ($M = 4.83$), and Scene Match ($M = 4.76$). Evaluators generally described the visuals as engaging and well-matched to the story content.

\begin{figure}[h]
    \centering
    \includegraphics[scale=0.4]{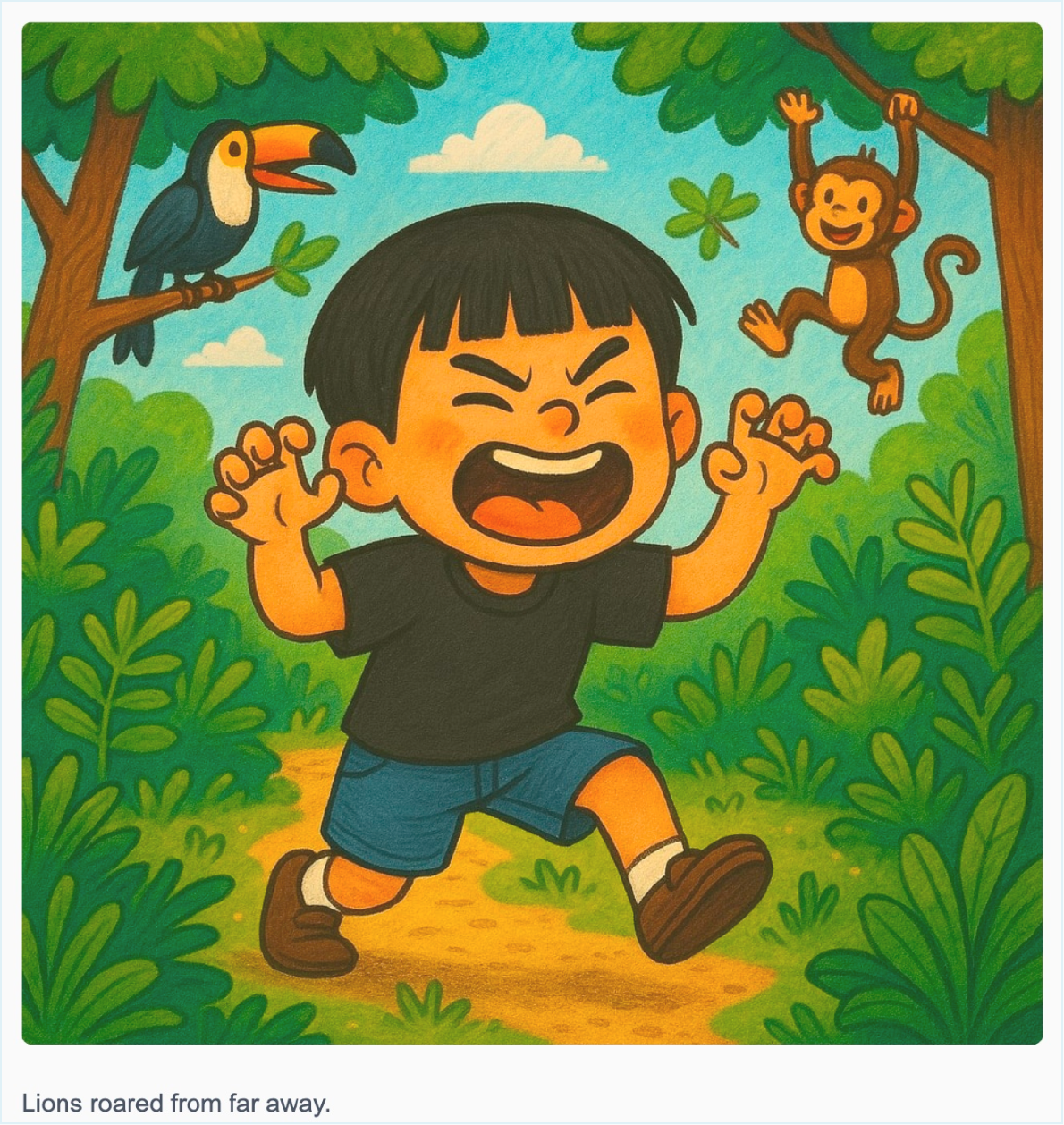}
    \caption{A scene from the ``Visit to the Zoo'' story depicting the zoo environment. Although the story text states that lions roared from far away, no lions are visible in the image, illustrating a mismatch between narrative content and visual representation.}
    \label{fig:zoo}
\end{figure}
\begin{figure}[h]
    \centering
    \includegraphics[scale=0.5]{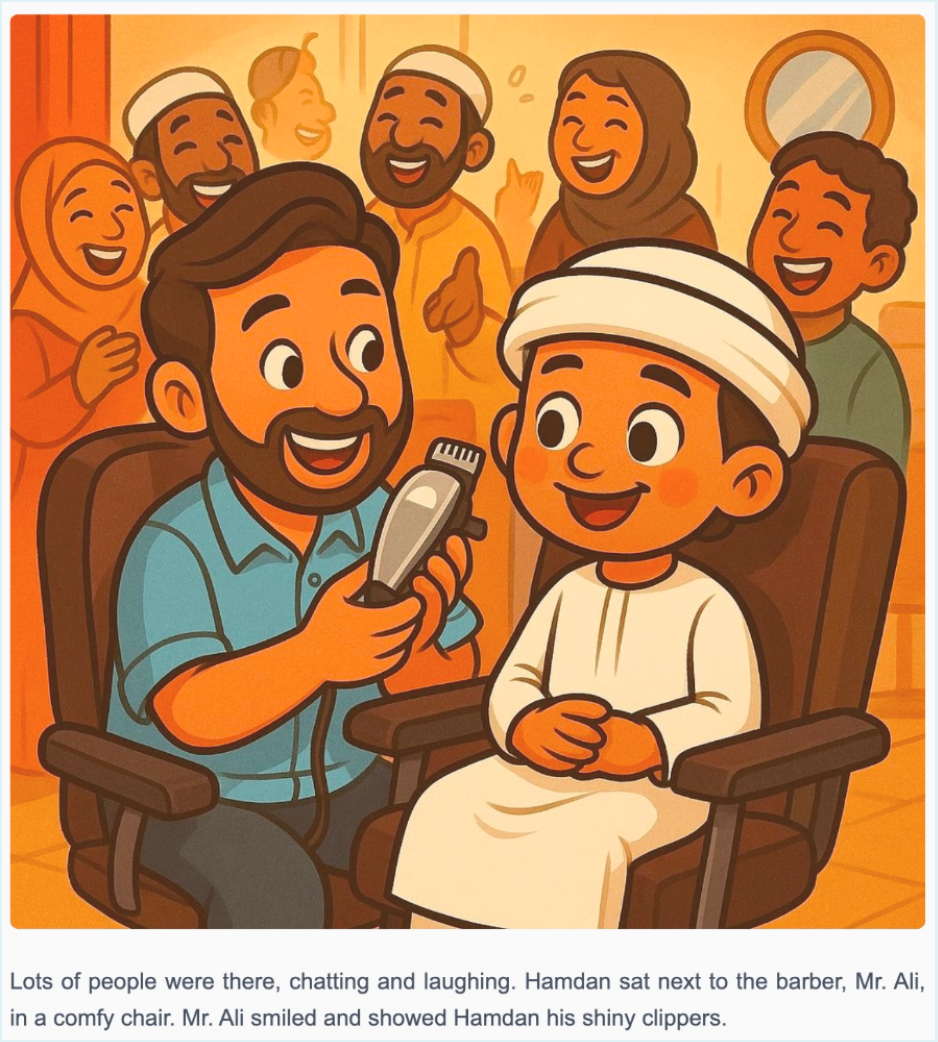}
    \caption{A scene from the ``Barbershop Visit'' story showing the protagonist seated with the barber. One evaluator noted that the background characters could be interpreted as mocking rather than casual socializing, which conflicts with the story’s intent to frame the barbershop as a welcoming environment.}
    \label{fig:barbershop}
\end{figure}
Open-ended responses identified a small number of scene-level issues. In one story, ``Visit to the Zoo,'' the text mentioned lions roaring from far away, but the corresponding scene visual (Figure~\ref{fig:zoo}) did not depict any lions, creating a mismatch between narrative content and imagery. Since Social Stories often rely on visuals to reinforce meaning, inconsistencies of this kind may weaken comprehension.

In another story, ``Barbershop Visit,'' the text described the barbershop as a busy but welcoming environment, with people chatting in the background. One evaluator noted that the generated visuals (Figure~\ref{fig:barbershop}) depicted background characters whose expressions and posture could be interpreted as laughing at or mocking the child protagonist, rather than simply socializing among themselves. The evaluator suggested replacing these characters with calmly seated figures, which would better match the story’s intent to portray the barbershop as a safe and reassuring setting.

Thus, these observations suggest that generated visuals were generally perceived positively, while also pointing to two recurring areas for improvement: maintaining semantic consistency between story text and scene imagery, and ensuring that background characters and environments reinforce rather than undermine the intended tone of the story.

\section{System Prompts}
\label{sec:prompts}
\subsection{Generating a New Social Story}
\label{sec:base-story-generation-prompt}

\begin{tcolorbox}[breakable,
    colback=white,
    arc=2mm,
    boxrule=0.5pt,
    top=2mm,
    bottom=2mm,
    left=2mm,
    right=2mm
]

Create a children's Social Story with the following specifications.

\medskip
\noindent\textbf{Core purpose}
\begin{itemize}
    \item Purpose: \texttt{\{purpose\}}
\end{itemize}

\noindent\textbf{Story details}
\begin{itemize}
    \item Target age: \texttt{\{age\_group\}} years old (default 6-8)
    \item Category: \texttt{\{category\}} (default "emotions")
    \item Word count: approximately \texttt{\{word\_count\}} words (default 200)
    \item Special guidelines: \texttt{\{guidelines\}} (optional)
\end{itemize}

\noindent\textbf{Writing instructions}
\begin{enumerate}
    \item Focus on helping children understand and cope with the specified situation.
    \item Use simple, clear language appropriate for the target age group.
    \item Keep sentences short (about 6-10 words on average).
    \item Structure the story into 3-5 clear paragraphs (scenes).
    \item Include positive coping strategies and reassurance.
    \item Make the story relatable and comforting.
    \item Do not include comprehension questions or titles.
\end{enumerate}

\noindent\textbf{Output format}
\begin{itemize}
    \item Return only the story text.
    \item Separate paragraphs or scenes with exactly two newline characters (\texttt{\textbackslash n\textbackslash n}).
    \item Do not include scene labels or numbers.
    \item Do not include any additional commentary.
\end{itemize}
\end{tcolorbox}

\subsection{Personalizing Story Text}
\label{sec:personalize-story-text-prompt}

\begin{tcolorbox}[breakable,
    colback=white,
    arc=2mm,
    boxrule=0.5pt,
    top=2mm,
    bottom=2mm,
    left=2mm,
    right=2mm
]

You are an AI assistant specialized in rewriting children's stories for children diagnosed with autism. Your goal is to personalize stories based on a student's profile so that they are able to relate to and comprehend the story better. You must maintain the core plot, moral, and tone of the original story. You must only return the rewritten story text, without any conversational filler, titles, or additional notes.

\medskip
\noindent\textbf{Student profile}
\begin{itemize}
    \item Student's first name: \texttt{\{first\_name\}}
    \item Comprehension level: \texttt{\{comprehension\_level\}} (or \texttt{"not specified"} if unknown)
    \item Preferred sentence length: \texttt{\{preferred\_sentence\_length\}} (or \texttt{"not specified"} if unknown)
    \item Preferred story length: \texttt{\{preferred\_story\_length\}} (or \texttt{"not specified"} if unknown)
    \item Learning preferences: \texttt{\{learning\_preferences\}} (or \texttt{"not specified"} if unknown)
    \item Challenges: \texttt{\{challenges\}} (or \texttt{"not specified"} if unknown)
    \item Additional notes: \texttt{\{additional\_notes\}} (optional)
\end{itemize}

\noindent\textbf{Original story}
\begin{itemize}
    \item Original story title: \texttt{\{original\_story\_title\}} (or \texttt{"Untitled Story"} if unknown)
    \item Original story content: \texttt{\{original\_story\_text\}}
\end{itemize}

\noindent\textbf{Instructions for rewrite}
\begin{itemize}
    \item Integrate only the student's first name \texttt{\{first\_name\}} naturally into the story.
    \item Adjust vocabulary and sentence structure to fit the student's comprehension level \texttt{\{comprehension\_level\}} (or a general children's level if not specified).
    \item Keep sentence lengths within the preferred range \texttt{\{preferred\_sentence\_length\}}:
    \begin{itemize}
        \item \texttt{very\_short}: 1 to 5 words
        \item \texttt{short}: 6 to 10 words
        \item \texttt{medium}: 11 to 15 words
        \item \texttt{long}: more than 15 words
    \end{itemize}
    \item Keep story length within the preferred range \texttt{\{preferred\_story\_length\}}:
    \begin{itemize}
        \item \texttt{very\_short}: 1 to 3 sentences
        \item \texttt{short}: 1 to 2 paragraphs
        \item \texttt{medium}: 3 to 5 paragraphs
        \item \texttt{long}: more than 5 paragraphs
    \end{itemize}
    \item Incorporate the student's interests and learning preferences naturally.
    \item Gently address the student's challenges if relevant.
    \item Maintain the core plot, moral, and a positive tone.
\end{itemize}

Only output the rewritten story text (no extra explanation, titles, or notes).
\end{tcolorbox}

\subsection{Generating Comprehension Check Activities}
\label{sec:comprehension-check-prompt}

\begin{tcolorbox}[breakable,
    colback=white,
    arc=2mm,
    boxrule=0.5pt,
    top=2mm,
    bottom=2mm,
    left=2mm,
    right=2mm
]

Based on the following Social Story, suggest a simple, real life, in the moment activity that a child with autism can do with their caregiver immediately after the story to practise the target skill.

\noindent\textbf{Student profile}
\begin{itemize}
    \item Student's first name: \texttt{\{first\_name\}}
    \item Comprehension level: \texttt{\{comprehension\_level\}} (or \texttt{"not specified"} if unknown)
    \item Story title: \texttt{\{story\_title\}}
    \item Story content: \texttt{\{story\_content\}}
    \item Interests: \texttt{\{interests\}} (or \texttt{"not specified"} if unknown)
\end{itemize}

\noindent The activity should be:
\begin{itemize}
    \item Concrete and doable in 2-5 minutes.
    \item Directly related to the core concept of the story.
    \item Does not require any additional materials or preparation.
    \item Suitable for the child's comprehension level and interests.
\end{itemize}

Only output the activity description (no extra explanation).
\end{tcolorbox}









\section{Mock-ups Used in Phase~1}
\label{sec:appendix-mockups}
\begin{figure*}
    \centering
    \begin{subfigure}[b]{0.49\textwidth}
        \centering
         \captionsetup{justification=raggedright, singlelinecheck=false, labelfont=bf, labelsep=period, font=small} 
        \caption{}
        \includegraphics[width=\textwidth]{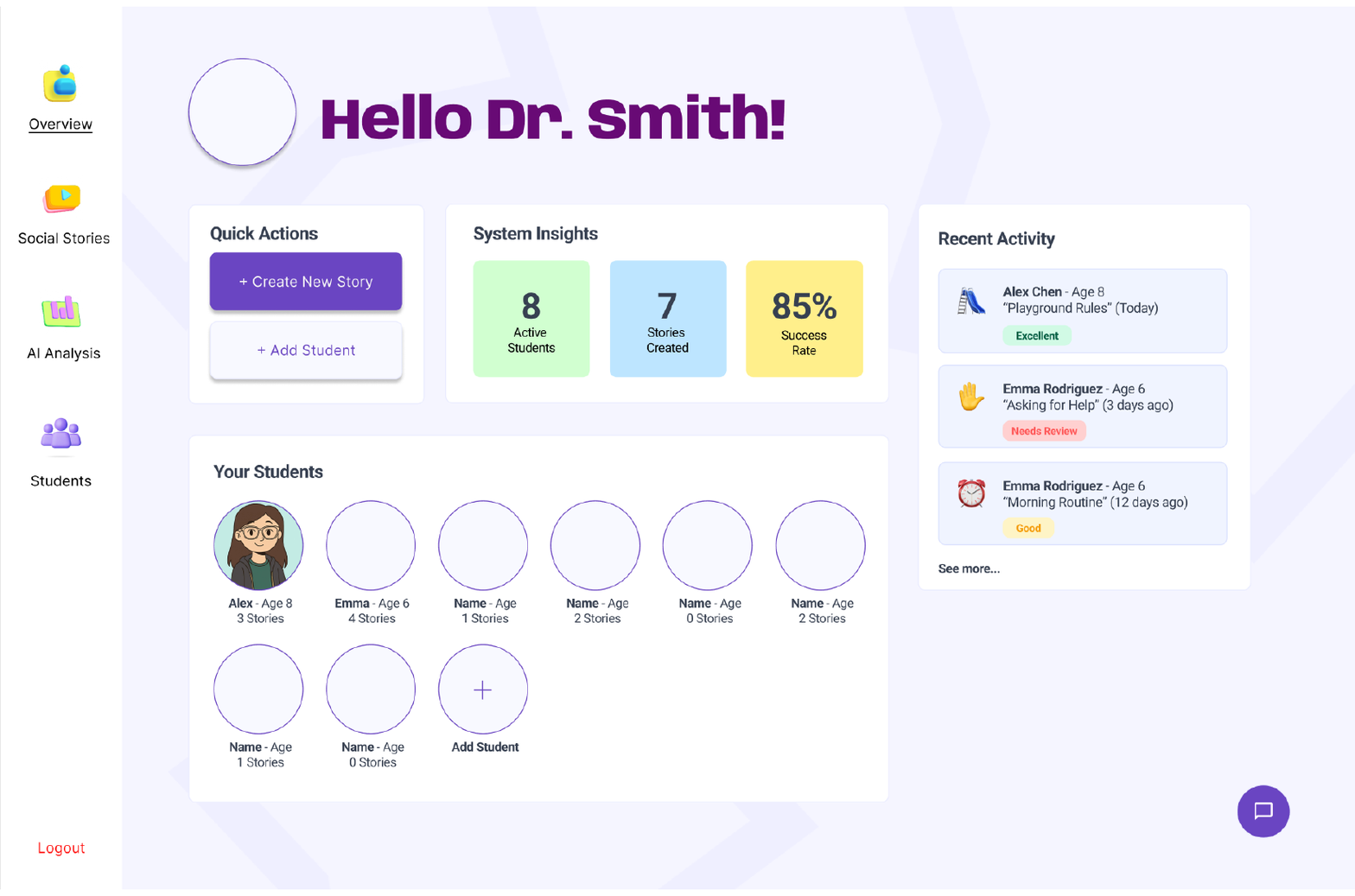} 
        \label{fig:overview-mockups}
    \end{subfigure}
    \hfill
    \begin{subfigure}[b]{0.49\textwidth}
        \centering
         \captionsetup{justification=raggedright, singlelinecheck=false, labelfont=bf, labelsep=period, font=small} 
        \caption{}
        \includegraphics[width=\textwidth]{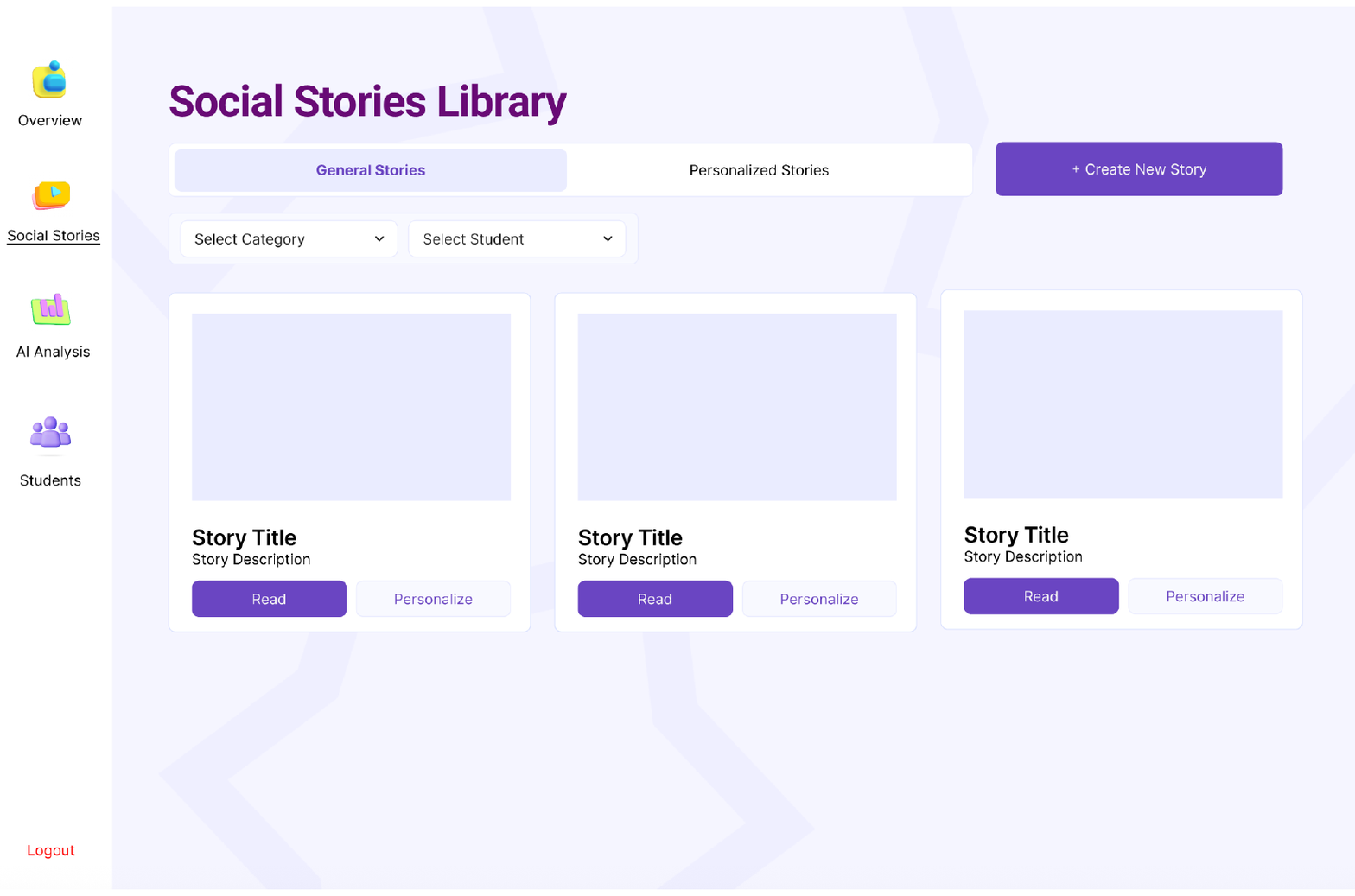} 
        \label{fig:social-story-mockup}
    \end{subfigure}

    \vspace{1em} 

    \begin{subfigure}[b]{0.49\textwidth}
        \centering
         \captionsetup{justification=raggedright, singlelinecheck=false, labelfont=bf, labelsep=period, font=small} 
        \caption{}
        \includegraphics[width=\textwidth]{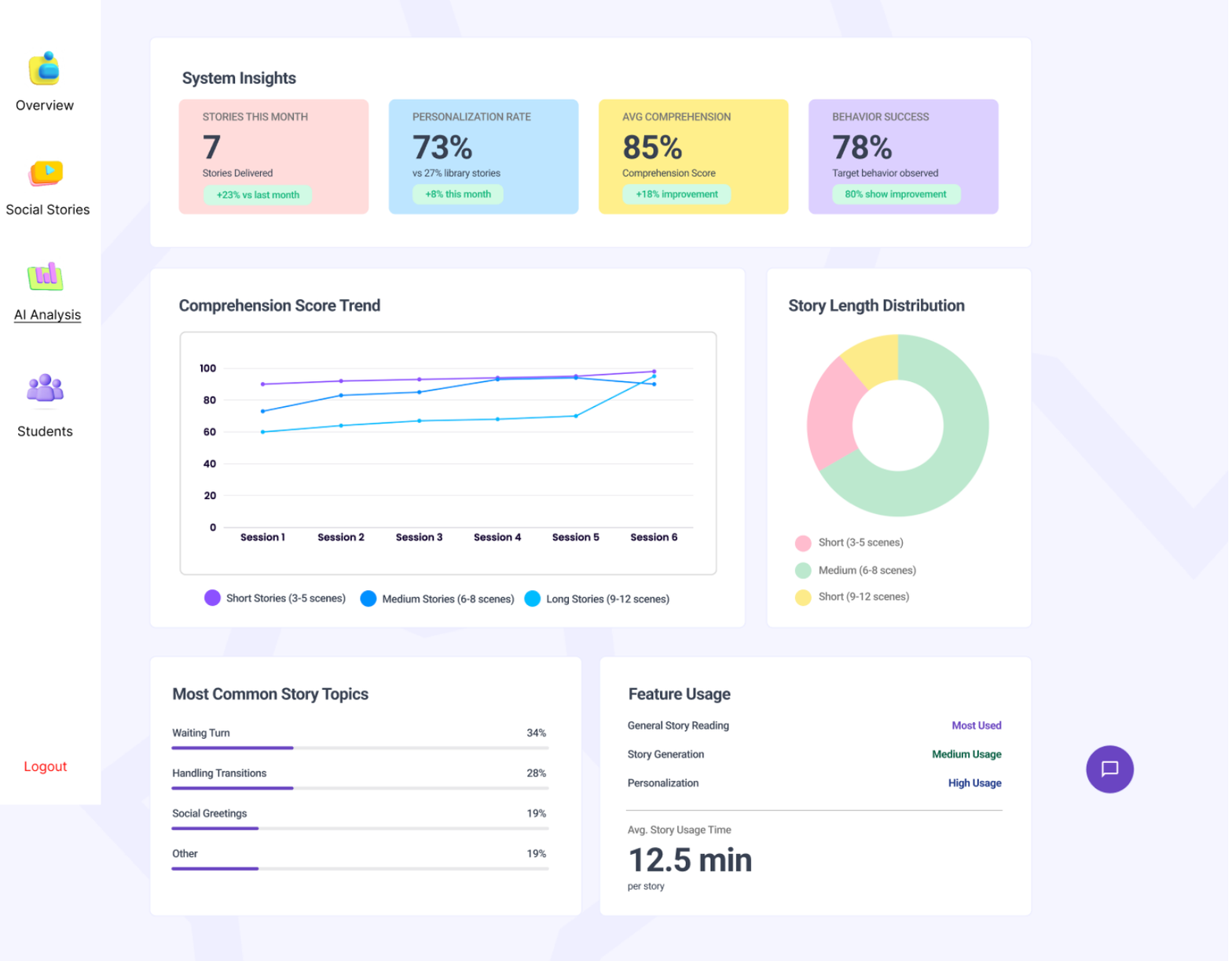} 
        \label{fig:ai-analysis-mockup}
    \end{subfigure}
    \hfill
    \begin{subfigure}[b]{0.49\textwidth}
        \centering
         \captionsetup{justification=raggedright, singlelinecheck=false, labelfont=bf, labelsep=period, font=small} 
        \caption{}
        \includegraphics[width=\textwidth]{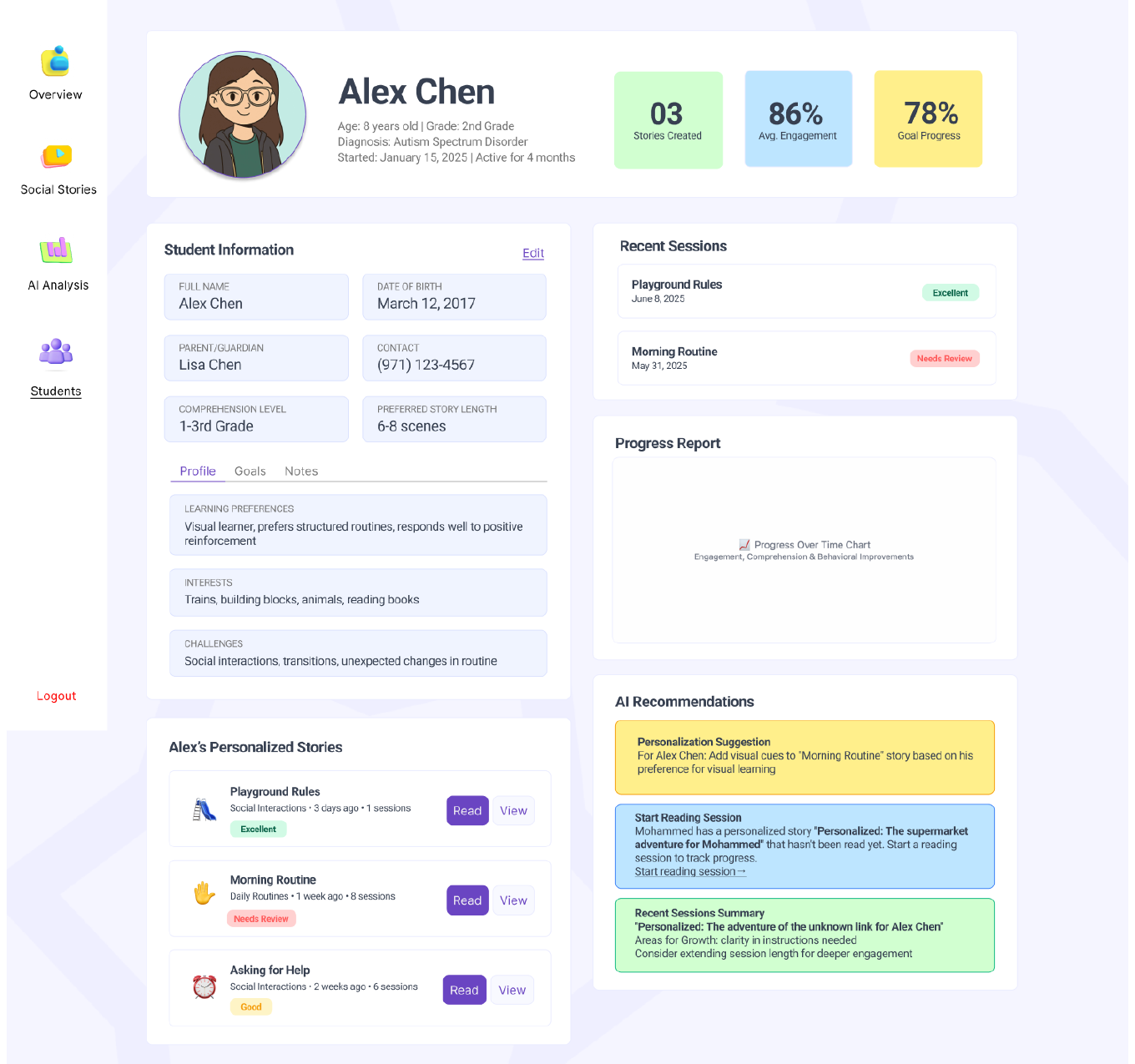} 
        \label{fig:students-mockup}
    \end{subfigure}

    \caption{Low-fidelity mock-ups from Phase 1. (A) Overview tab showing quick actions and recent activity for the practitioner. (B) Social Stories tab with access to the system library, personalized stories, and the option to create stories using AI. (C) Analysis tab showing student-wide aggregate patterns and system usage analytics, including comprehension trends and story use. (D) Students tab showing an individual student profile with key information, personalized stories, progress summaries, and AI-generated recommendations.}
    \label{fig:mockups}
\end{figure*}

\appendix
\end{document}